\documentclass[aps,twocolumn,amsmath,amssymb]{revtex4}
\pdfoutput=1
\usepackage{graphic x}
\usepackage{bm}

\usepackage{hyperref}

\def\cal#1{\mathcal{#1}}
\def\beq{\begin{equation}}
\def\eeq{\end{equation}}
\def\bea{\begin{eqnarray}}
\def\eea{\end{eqnarray}}

\def\en{n^{\star}}

\def\kt{k_{\rm B} T}
\def\kb{k_{\rm B}}

\begin{document}

\title{The role of collective motion in examples of coarsening and self-assembly}

\author{Stephen Whitelam$^{1,2,3}$} 
\author{Edward H. Feng$^2$} 
\author{Michael F. Hagan$^4$} 
\author{Phillip L. Geissler$^{2,3}$}
\affiliation{$^1$Systems Biology Centre, University of Warwick, Coventry CV4 7AL, UK\\
$^2$Department of Chemistry, University of California at Berkeley, Berkeley, CA 94720, USA\\ $^3$Physical Biosciences and Materials Sciences Divisions, Lawrence Berkeley National Laboratory, Berkeley, CA 94720, USA\\
$^4$ Department of Physics, Brandeis University, Waltham, MA, USA}

\date{\today}

\begin{abstract}
The simplest prescription for building a patterned structure from its constituents is to add particles, one at a time, to an appropriate template. However, self-organizing molecular and colloidal systems in nature can evolve in much more hierarchical ways. Specifically, constituents (or clusters of constituents) may aggregate to form clusters (or clusters of clusters) that serve as building blocks for later stages of assembly. Here we evaluate the character and consequences of such collective motion in a set of prototypical assembly processes. We do so using computer simulations in which a system's capacity for hierarchical dynamics can be controlled systematically. By explicitly allowing or suppressing collective motion, we quantify its effects. We find that coarsening within a two dimensional attractive lattice gas (and an analogous off-lattice model in three dimensions) is naturally dominated by collective motion over a broad range of temperatures and densities. Under such circumstances, cluster mobility inhibits the development of uniform coexisting phases, especially when macroscopic segregation is strongly favored by thermodynamics. By contrast, the assembly of model viral capsids is not frustrated but is instead facilitated by collective moves, which promote the orderly binding of intermediates consisting of several monomers.
\end{abstract}

\maketitle
\section{Introduction: self-assembly and collective motion} 

Self-assembly refers to the generation of patterns or aggregates through the interaction of autonomous components~\cite{whitesides}. The supramolecular self-assembly of thermodynamically stable structures plays a central role in biology, notably in the hybridization of nucleic acid strands, the organization of lipids to form cell membranes, and the assembly of proteins to form casings for viruses and other genetic material~\cite{virus_ref}. Biological components can also self-assemble when removed from their natural environment: protein complexes called chaperonins, for example, form large-scale sheet-like~\cite{chap} and string-like~\cite{chap_cyto} structures  {\em in vitro}. Self-assembly is widespread in non-biological contexts, from the myriad patterns formed by soap and oil in water to the ribbon-like structures that assemble from cobalt nanoparticles in solution~\cite{cobalt}. The formation of such structures inspires the design of materials with novel properties, the chief goal of nanotechnology.
\begin{centering}
\begin{figure}[h!]
\includegraphics[width=8cm]{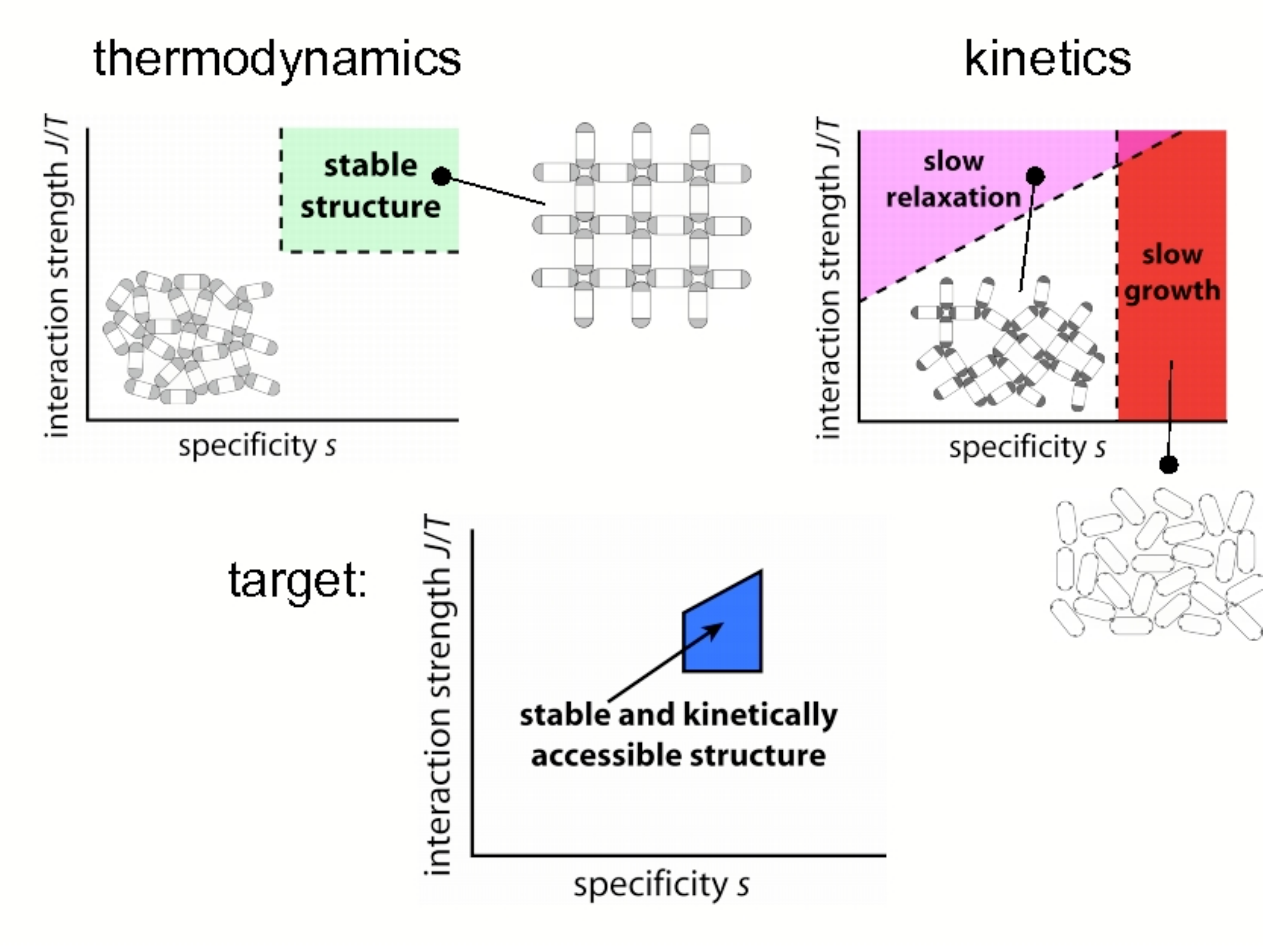} 
\caption{\label{figcomp} The competitive nature of self-assembly: an illustration of the antagonism between the requirements for structural stability and kinetic accessibility. We consider a schematic system of ovoid nanoparticles (mimicking, for example, inorganic nanorods), equipped with pairwise end-to-end interactions of strength $J$ (darker shading of rod ends denotes greater interaction strength) and specificity, or inverse angular tolerance, $s$ (inversely proportional to shaded area). Stable structures of the required symmetry in general require strong, specific interactions (top left). However, such requirements tend to frustrate the kinetics of assembly (top right): too strong an interaction and structures fail to relax as they grow; too specific an interaction and productive binding events are rare. The competition between kinetics and thermodynamics dictates the region of viable assembly (bottom).}
\end{figure}
\end{centering}

The success of self-assembly is determined by an interplay of thermodynamics and dynamics. Components equipped with strong and highly directional interactions may stabilize the thermodynamically preferred structure, but are not guaranteed to assemble spontaneously into such a structure. Overly strong attractions may prevent structural relaxation (by impairing unbinding~\cite{key_lock,break2} or `reversibility'~ \cite{new_rapaport}), resulting in malformed aggregates. Components that must bind via precise alignment with a neighbor may take a prohibitively long time to do so, resulting in slow structural growth. This competition between the requirements for structural stability and kinetic accessibility in general restricts viable assembly to small regions of parameter space, an idea illustrated in Figure~\ref{figcomp}.

Identifying and controlling regimes of viable assembly pushes the envelope of current experimental capabilities. In support of such work, computer simulation provides a powerful means of understanding self-assembly and its potential for creating new materials. Simulation permits exhaustive trials of model systems at little expense~\cite{mike,Glotzer0,Glotzer1,patchy1,patchy2}. It can reveal both the nature of inter-component forces that lead to thermodynamically stable structures, and the dynamics through which such components associate~\cite{key_lock,mike,new_rapaport}. When assessing the assembly properties of a given model it is desirable to evolve that model in order to approximate the dynamics that the corresponding physical system would execute. Molecular dynamics~\cite{frenkel_book,art} algorithms evolve components according to Newton's laws of motion, and so are a natural choice for simulating particle systems. However, self-assembling components generally possess anisotropic interactions of maximal strength much greater than $\kb T$ and range much less than a particle diameter. Strong, short-ranged interactions place stringent limits on the maximum integration time step able to preserve numerical stability. Under such conditions, simulations of large-scale assemblies are very time-consuming, forcing the simulator to choose between focusing on dynamics on smaller scales or starting from forcibly equilibrated samples.

One way to circumvent this problem is to use a coarse-grained dynamical procedure to move particles according to potential energy gradients without explicitly integrating equations of motion. The Monte Carlo technique provides a flexible framework in which to do so~\cite{frenkel_book,mc_review,mc_dynamics1,mc_dynamics2,mc_dynamics3,mc_dynamics4}.  However, conventional Monte Carlo techniques involve sequential moves of individual particles and so neglect the correlated motion of particles on timescales less than the fundamental discrete time step $\Delta t$ (the time corresponding to a typical discrete particle displacement). For some systems this neglect of collective motion appears to be unimportant: in Ref. ~\cite{mc_dynamics4}, for example, molecular dynamics results were reproduced using a single-particle Monte Carlo protocol. For many systems, however, chiefly those whose constituents possess interactions whose strengths vary strongly with angle or distance, neglecting motion correlated on timescales less than $\Delta t$ leads to unphysical relaxation, particularly at long times.

To address such problems, `cluster' algorithms have been used extensively to effect correlated or collective motion~\cite{SW, big_bad_wolff, wu,panag_cluster, amar, Liu1, mak, babu,krauth}. In general, such algorithms identify collections of particles to be moved in concert by recursively `linking' particles according to a set of criteria, such as the pairwise energy or degree of proximity of neighboring particles. One such algorithm, the `virtual-move' Monte Carlo (VMMC) procedure of Ref.~\cite{vmmc}, is designed to effect correlated displacements and rotations according to potential energy gradients or forces experienced under `virtual' moves of neighboring particles. We describe this idea in the following section and in the Appendix. This procedure reduces to single-particle motion when particles experience small energy changes on timescale $\Delta t$, but effects collective motion on arbitrarily large lengthscales when particles experience large changes in energy on this timescale.

In this paper we use VMMC to study the qualitative effect of collective motion on self-assembly by explicitly allowing or suppressing collective moves of particles. In Section~\ref{section_lattice} we consider the 2$d$ attractive lattice gas, which coarsens upon a temperature quench via the self-assembly of the homogeneous phase. We find a range of temperatures at which inter-particle forces are large enough to encourage assembly but not so large that motion is strongly correlated on timescales less than $\Delta t$. In this case assembly is driven by single-particle binding and unbinding events, and little qualitative effect is observed upon accounting explicitly for correlated motion. However, at low temperatures we find that motion on a timescale $\Delta t$ is strongly correlated according to potential energy gradients. Neglecting or allowing explicit collective motion under these conditions selects drastically different fates for the system: strongly collective motion impairs assembly via the formation of kinetic traps associated with the binding of large clusters. In the language of coarsening, these kinetic traps represent the arrest of phase separation by gelation. Suppressing collective motion suppresses gelation. We present a simple argument designed to estimate the importance of collective motion in the space of temperature and particle concentration.

In Section~\ref{sec_colloid} we examine a three-dimensional off-lattice system of hard spheres with isotropic pairwise square-well interactions, a model of strongly-associating colloids. Here, as for the lattice gas, collective motion tends only to impair assembly (or promote gelation) by inducing awkward binding events between large clusters. However, in Section~\ref{sec_capsid}, we study a model of viral capsid assembly that displays qualitatively different behavior. Inter-particle forces are sufficiently strong that substantial correlated motion emerges on the fundamental timescale $\Delta t$, but the geometry of inter-particle association is such that these collective motions {\em improve} assembly by inducing productive collisions between small intermediates larger than monomers. Suppressing this correlated motion slows assembly, but does not strongly impact the final capsid yield at most thermodynamic states. Collective motion therefore appears to play a qualitatively different role within different model systems: it drives the formation of kinetic traps {\em and} facilitates orderly growth.

\section{A `virtual-move' Monte Carlo Algorithm}
\label{section_algorithm}

\begin{centering}
\begin{figure}[h!]
\includegraphics[width=8cm]{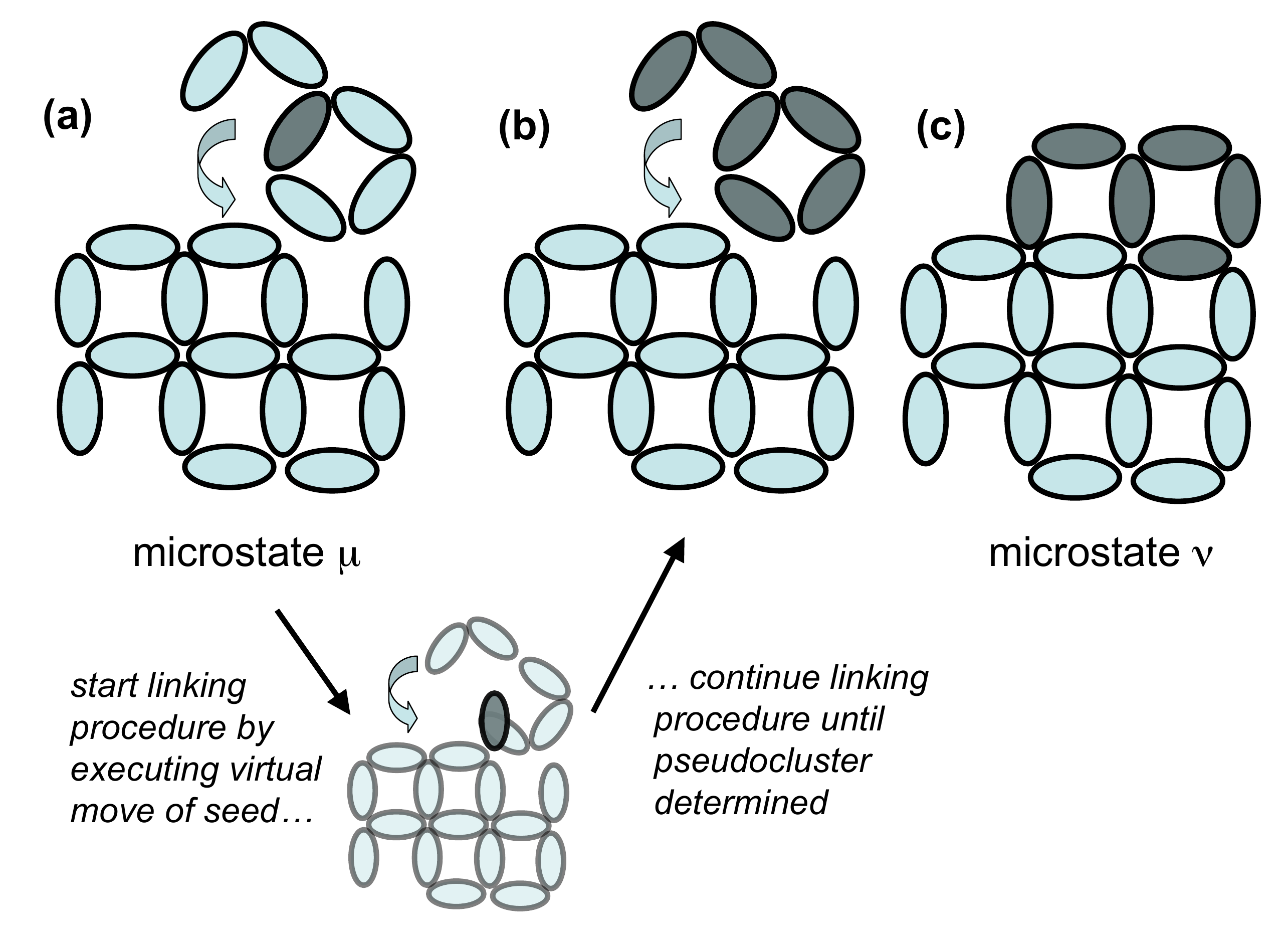} 
\caption{\label{figvmmc} Illustration of the VMMC procedure applied to a collection of pairwise-interacting oval nanoparticles. (a) Starting in microstate $\mu$ we pick a seed particle (shaded) from which to `grow' a pseudocluster, together with a virtual move map (denoted by the arrow). The pseudocluster is a set of linked particles that will experience a trial move. The virtual move map is both a device for computing neighboring potential energy gradients, and defines the move that the pseudocluster will execute. The pseudocluster growth procedure is an iterative linking scheme akin to the Swendsen-Wang (SW) algorithm~\cite{SW}: we propose links between the seed and all particles with which it interacts, and continue iteratively until links have been proposed between the pseudocluster and all particles with which its constituents interact. Our procedure differs from the SW algorithm in that here links are formed not on the basis of pairwise energies, but instead on the basis of pairwise energy {\em gradients}. These gradients are computed by executing virtual moves of shaded particles (example shown in lower panel), and recording neighboring pairwise energies before and after those moves. Conditioning the acceptance criterion upon reverse virtual moves (not shown) ensures that detailed balance is preserved. The linking criterion is similar to that of the geometric cluster algorithm of Liu and Luijten~\cite{Liu1}, although here we propose links only between particles that interact in the {\em initial} configuration. We seek to effect dynamically realistic local and collective motion, rather than the collective and {\em nonlocal} equilibration-speeding motion of Ref.~\cite{Liu1}. With the pseudocluster so defined (b) we displace it according to the virtual move map, resulting in microstate $\nu$ (c). We then evaluate the Monte Carlo acceptance criterion (see text) and accept or reject the move.}
\end{figure}
\end{centering}
Here we summarize the virtual-move Monte Carlo algorithm. We consider a $d$-dimensional collection of $N$ particles equipped with pairwise interactions. This algorithm is a dynamic procedure designed to identify, on the basis of potential energy gradients explored on a fundamental timescale $\Delta t$, the extent to which the motion of one particle is correlated with that of its neighbors. If so correlated, this motion is effected with a frequency designed to approximate a physical dynamics. An illustration of the procedure is given in Figure~\ref{figvmmc}. We start in microstate $\mu$. We define a `pseudocluster' ${\cal C}$ (a set of particles to be moved in concert) by choosing as its first member a `seed' particle $i$. We link the seed to a neighbor $j$ with probability $p_{ij}(\mu \to \nu)$, which in general depends on a `virtual' move of $i$ that defines a notional new microstate, $\nu$. Particles linked to members of the pseudocluster join the pseudocluster. We proceed iteratively, until no more members are added to the pseudocluster. We accept the move $\mu \to \nu$ with probability
\bea
\label{factor3}
&\,&W_{\rm acc}^{\left(\mu \to \nu|{\cal R}\right)} \nonumber \\
&=& \Theta \left(n_{\rm c} -n_{{\cal C}} \right) \cal{D}(\cal{C}) \, \min \left\{ 1, {\rm e}^{-\beta(E_{\nu}-E_{\mu})} \nonumber \right. \\ 
&\times& \left. \frac{\prod_{\nu \to \mu} q_{ij}(\nu \to \mu)}{\prod_{\mu \to \nu} q_{ij}(\mu \to \nu) } \prod^{{\cal R}}_{\langle i j \rangle_{\ell}} \frac{p_{ij}( \nu \to \mu)}{p_{ij}( \mu \to \nu)} \right\} .
\eea
Here $\cal{D}({\cal C}) \leq 1$ is a factor we impose in order to modulate the diffusivity of pseudoclusters according to size; $E_{\alpha}$ is the energy of the system in microstate $\alpha$; $q_{ij} \equiv 1-p_{ij}$ is the probability of not linking particles $i$ and $j$; and ${\cal R}$ denotes a particular realization of formed and failed links. The link-forming procedure is aborted {\em in situ} if the pseudocluster size $n_{{\cal C}}$ exceeds a specified cutoff $n_{{\rm c}}$, the smallest integer larger than $\xi ^{-1}$. Here $\xi $ is a random variable drawn uniformly from the interval $[0,1]$. The subsequent rejection of the move is enforced by the factor $\Theta(n_{\rm c} - n_{{\cal C}})$. This rejection procedure ensures that all particles experience proposed moves with approximately equal frequency. The products over $q$ variables quantify the probabilities of not forming proposed links, internal to and external to ${\cal C}$ (for forward and reverse moves), while the products over $p$ variables quantify the probabilities of linking together members of the pseudocluster. We choose to link particles $i$ and $j$ with a probability
\bea
\label{virtual_link}
p_{ij}(\mu \to \nu)&=&\Theta \left(n_{\rm c} -n_{{\cal C}}  \right) \nonumber \\
&\times& \cal{I}_{ij}^{(\mu)} \textnormal{max}\left(0,1-{\rm e}^{\beta \epsilon(i,j)-\beta \epsilon(i',j)}\right)
\eea
that depends on a virtual move (translation or rotation) of $i$ relative to $j$. Here $\epsilon(i,j)$ is the pairwise energy of the bond $ij$ in microstate $\mu$, and  $\epsilon(i',j)$ is the bond energy following the virtual move of $i$. The factor $\cal{I}_{ij}^{(\mu)}$ is unity if $i$ and $j$ interact in microstate $\mu$, and zero otherwise; the factor $\Theta \left(n_{\rm c} -n_{{\cal C}}  \right)$ terminates link formation if $n_{{\rm c}} > n_{{\cal C}}$. Linking particles in this fashion ensures that neighbors exert mutual forces proportional to the gradient of their pairwise energies, with motion `linked' or correlated if $\beta \epsilon(i',j) -\beta \epsilon(i,j)$ is large. Equation~(\ref{virtual_link}) implies that the acceptance rate~(\ref{factor3}) reduces to
\bea
\label{accept}
&\,&W_{\rm acc}^{(\mu \to \nu|\cal{R})} = \nonumber \\
&\,& \Theta \left(n_{\rm c} -n_{{\cal C}} \right) \cal{D}(\cal{C}) \min \left\{1, \prod_{\langle i j \rangle_{{\rm n}\leftrightarrow {\rm o}}} e^{-\beta\left( \epsilon_{ij}^{(\nu)} -\epsilon_{ij}^{(\mu)} \right)}  \right. \nonumber \\ 
&\times& \left. \prod_{\langle i j \rangle_{{\rm f}}} \frac{ q_{ij}( \nu \to \mu)}{q_{ij}( \mu \to \nu)}  \prod^{{\cal R}}_{\langle i j \rangle_{\ell}} \frac{ p_{ij}( \nu \to \mu)}{p_{ij}( \mu \to \nu)} \right\}.
\eea
The label $\langle ij \rangle_{{\rm n} \leftrightarrow {\rm o}}$ identifies particle pairs that start ($\mu$) in a noninteracting configuration and end ($\nu$) with positive energy of interaction (overlapping), {\em or} which start ($\mu$) in an overlapping configuration and end ($\nu$) in a noninteracting one. The final line of Equation~(\ref{accept}) accounts explicitly for links $\langle i j \rangle_{\ell}$ and failed links $\langle i j \rangle_{\rm f}$ internal to the pseudocluster (note that the $q$ factors internal to the pseudocluster were omitted in error in Equations (11) and (13) of Ref.~\cite{vmmc}). In Appendix A we outline a procedure in which this acceptance rate is simplified by `symmetrizing' link formation at the level of link generation.

\section{A prototype of self-assembly: the attractive lattice gas}
\label{section_lattice}

In this section we consider the quench-driven coarsening of the 2$d$ attractive lattice gas with conserved particle number. Lattice gas models, whose thermodynamics can be related to the Ising model~\cite{chandler,huang}, are used to caricature a diverse range of physical systems, from binary metallic alloys to solvent-mediated nanoparticle aggregation~\cite{rabani}. We regard coarsening within the attractive lattice gas as a prototype of self-assembly in which attractive interactions drive the organization of a homogeneous phase. Our aim is to identify the range of temperatures and particle concentrations where collective motion (motion correlated on a timescale less than the discrete time step $\Delta t$) strongly influences assembly.

Many authors have studied the dynamics of the lattice gas using single-particle Monte Carlo algorithms, inducing transitions between microstates by moving a single particle to an unoccupied nearest-neighbor site. The assumption underlying these studies is that sequential moves of single particles represent a good approximation of the dynamics that the corresponding physical system might execute. This assumption is likely accurate for systems that relax via transport of mass from interfaces of high to low curvature by diffusion through the intervening medium~\cite{bray}. Scaling arguments and simulations based on this physical picture predict the typical domain size $L$ to grow as $L(t) \sim t^{1/3}$~\cite{bray,amar}.

However, in many settings, the transport of mass between domains by the evaporation and diffusion of constituent monomers is not the only possible mode of relaxation. Indeed, we frequently encounter the concerted motion of domains of one phase in another: witness the relaxation of polymers in solution or the the diffusion of nanocrystal aggregates on graphite~\cite{ge}. To model such behavior within the lattice gas, we must explicitly account for motion correlated on timescales of order $\Delta t$. 

We consider a collection of $N$ particles on a simple square lattice of $ V=L^2$ sites. Two particles may not occupy the same site, and interact with binding energy $-\epsilon_{\rm b}$ when occupying nearest-neighbor sites. We express temperature $T $ in units of $\epsilon _{ {\rm b} } / k _{ {\rm B} }$. Particles are dispersed randomly on the lattice with concentration $\phi_0 = N/V$ and are evolved using virtual-move Monte Carlo translations. We enforce a scaling of the diffusion constant of $n^{-\alpha}$ for clusters of size $n \geq 1$. We focus on the difference between the case $\alpha = \infty$, corresponding to single-particle moves, and $\alpha=1$, corresponding to diffusion akin to Brownian dynamics.

\begin{figure}[h]
\includegraphics[width= 8.5cm]{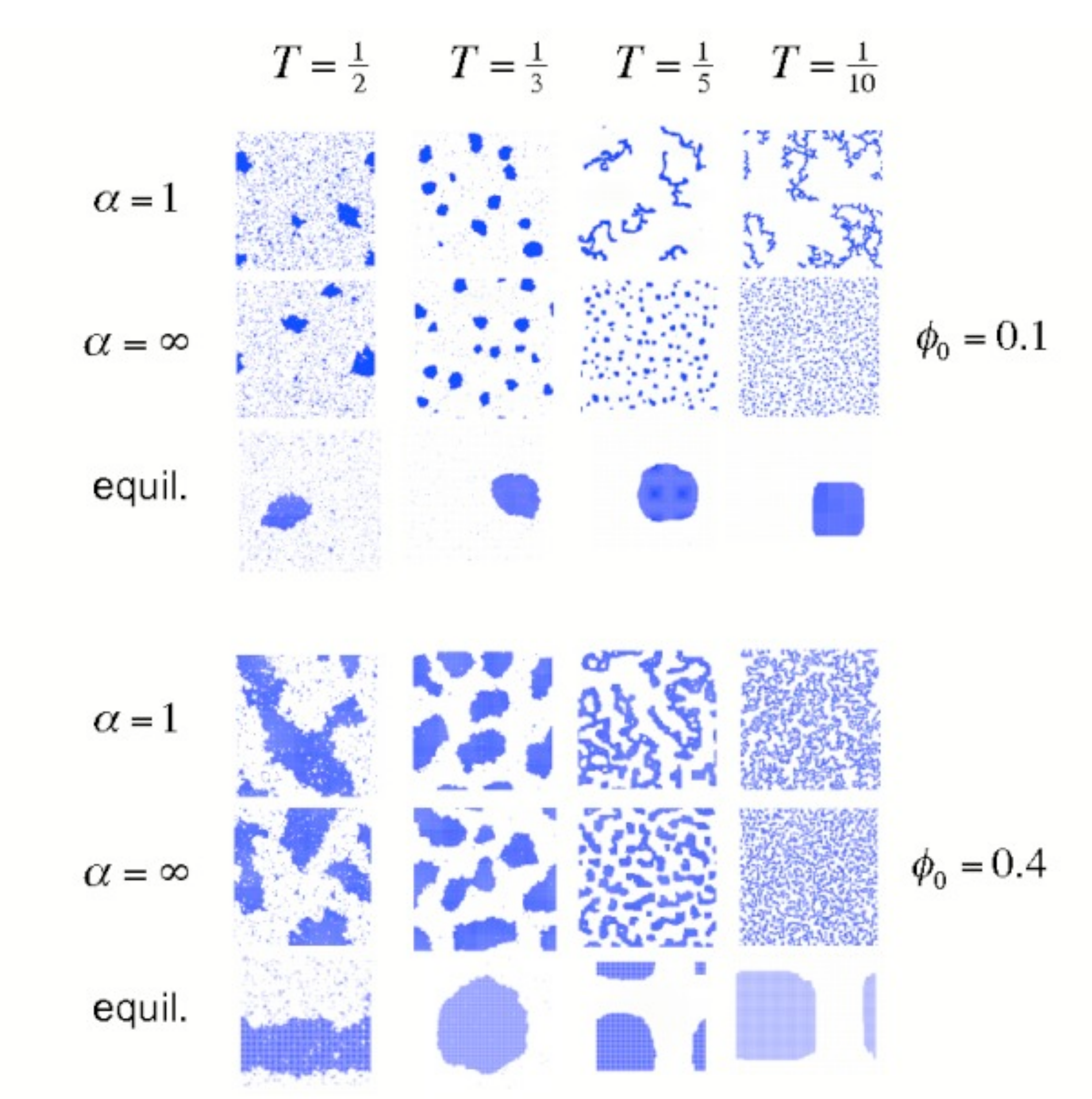} 
\caption{\label{latticepic} Configurations of the lattice gas at fixed time following a quench from a disordered state, evolved using Monte Carlo dynamics with cluster diffusivity $D(n) \propto n^{-\alpha}$ and particle concentrations $\phi_0=0.1$ (top panel) and 0.4 (bottom panel). The equilibrium states (obtained using nonlocal moves, panel `equil.') are in all cases phase-separated configurations. Top panel: At high temperature, collective ($\alpha=1$)- and single-particle ($\alpha=\infty$) motion give rise to visually similar assembly behavior, while at low temperature collective motion drives the formation of gel-like kinetically trapped structures. Gelation is avoided by suppressing collective motion. Similar behavior is seen at higher concentrations (bottom panel), although collective motion at the lower temperatures is hindered sterically. Coarsening images captured after 2 million MC steps per particle (1.5 million steps for $T=1/3$ and $\phi_0=0.4$).}
\end{figure}

The simplicity of the model allows us to estimate the values of temperature $T$ and particle concentration $\phi_0$ for which we expect motion correlated on a timescale $\Delta t$ to be important. We present this argument for the $d$-dimensional hypercubic lattice gas. We assume for simplicity that growing clusters are compact (as a two-dimensional illustration, consider the left half of Figure~\ref{latticepic}, top row). We wish to estimate the timescale on which two clusters of size $n$ encounter each other as a result of their collective diffusion, $\tau_{\rm enc}(n,n)$, and compare this estimate with the timescale on which such clusters exchange mass via evaporation (Ostwald `ripening'~\cite{ripe_ref}), $\tau_{\rm evap}(n)$. If $\tau_{\rm evap}(n)$ exceeds $\tau_{\rm enc}(n)$, we expect collective modes of motion to significantly influence assembly.

Within the virtual-move algorithm the probability of whole-cluster motion is approximately
\beq
\label{whole}
p_{\rm whole}(n) =  p_{\rm link}^{(1+\xi)(n-1)} \frac{1}{n^{1+\alpha}},
\eeq
where $p_{\rm link} \equiv 1-e^{-\beta \Delta \epsilon}$ is the probability of linking two particles following a virtual move, and $\Delta \epsilon = \epsilon_{\rm b}>0$ is the change in energy resulting from separating those particles. We impose a factor of $n^{-1}$, which ensures that particles suffer attempted moves with approximately equal frequencies, and a factor of $n^{-\alpha}$, to account for our chosen diffusion constant. The number $\xi$ reflects the efficiency with which the recursive algorithm forms links within a cluster. This number is approximately $1/(2d)$ for the version of the algorithm described in the main text, and is zero for the version of the algorithm described in the appendix. This distinction is unimportant at low temperature, and we shall for simplicity take $\xi =0$. 

The probability that a single particle breaks away from its host cluster is
\beq
\label{unbind}
p_{\rm unbind}(n) \approx  e^{-\beta z_n \Delta \epsilon} {\cal G}_n,
\eeq
where 
\beq
{\cal G}_n = 2d \frac{n^{(d-1)/d}}{n} 
\eeq
is a geometric factor quantifying the likelihood that a chosen particle lies on the surface of a cluster (we expect this approximation to be reasonable for $n >10$), and $z_n$ is the typical coordination number of a particle on the surface of a cluster of size $n$.
\begin{center}
\begin{figure}[ht]
\includegraphics[width=7cm]{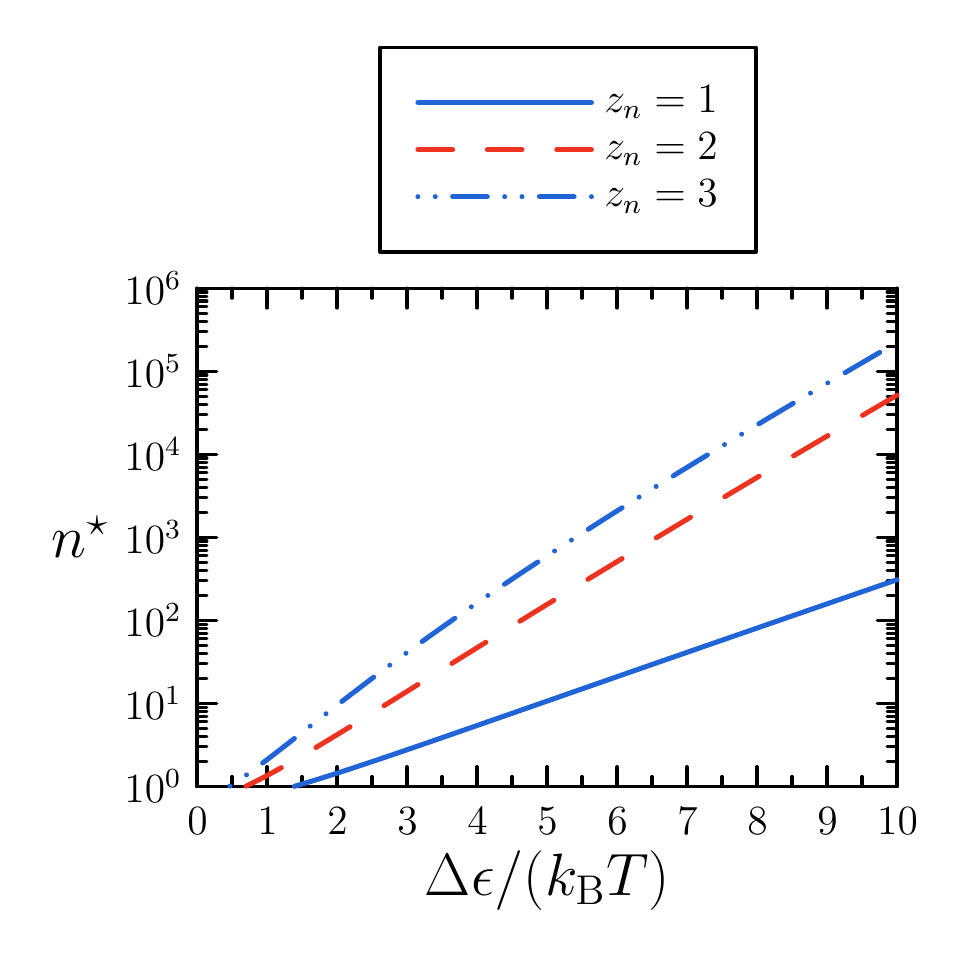} 
\caption{\label{figclus1} Estimate of the size $\en$ of the largest cluster for which collective motion is at least as probable as the unbinding of its constituent monomers, as a function of particle binding energy $\Delta \epsilon$. Estimate is derived from a simple argument (Equation~(\ref{clustodal})) pertaining to the 2$d$ lattice gas, assuming constant coordination number $z_n=1, 2$ and 3. For the two larger values of $z_n$, $\en$ is appreciable even for relatively modest values of $\Delta \epsilon$.}
\end{figure}
\end{center}
In $d=2$ we have that $p_{\rm whole}(n) \propto {\rm e}^{-(n-1) | \ln p_{\rm link}|} n^{-1-\alpha}$, while $p_{\rm unbind}(n) \propto n^{-1/2}$. For sufficiently large $n$ we therefore expect single-particle binding events to dominate. To determine the cluster size $\en$ at which unbinding and collective motion are equally likely, we equate Equations~(\ref{whole}) and~(\ref{unbind}):
\beq
\label{clustodal}
(\en-1) \ln p_{\rm link} =- \beta z_{\en}\, \Delta \epsilon + \ln (2d)+\left(\frac{d-1}{d} + \alpha \right) \ln \en.
\eeq
Clusters smaller than $\en $ will likely move as a whole.  To estimate
 $\en $ at low temperatures, we ignore the term logarithmic in $\en$ to obtain 
\beq
\label{lowt}
\en(T) \sim \frac{\Delta \epsilon  }{k_{\rm B} T} z_{\en} \,e^{\Delta \epsilon/(k_{\rm B} T) }.
\eeq
In Figure~\ref{figclus1} we plot $\en$ obtained from numerical solution of Equation~(\ref{clustodal}) for $d=2$, $\alpha=1$, and three values of $z_n$. We observe that $\en$ is large even for relatively modest values of $\Delta \epsilon/(\kt)$ when $z_n\geq2$.
\begin{figure}[ht]
\includegraphics[width=8cm]{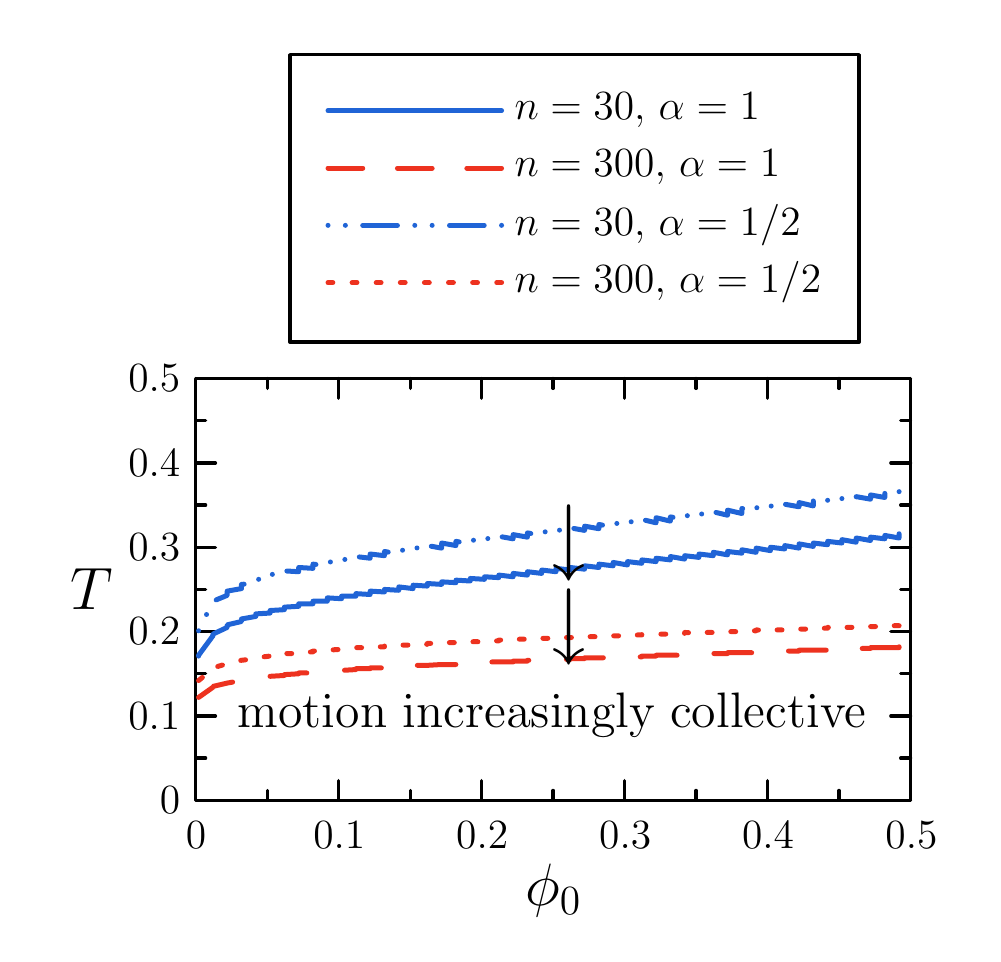} 
\caption{\label{figclus2} Estimate of the regime of collective motion for the 2$d$ attractive lattice gas for the given values of cluster size $n $ and diffusion exponent $\alpha $. We show the intersection of the `clustodal' (locus of ${\cal P}_{\rm coll}(n)=1$) with surfaces of constant cluster size $n$ as a function of $T$ and $\phi_0$.  For given $n$ and $\alpha$, collective motion dominates the behavior of clusters of size $n$ (or smaller) in the region below the corresponding line. Hence, collective motion dominates this kinetic phase diagram for a broad range of temperatures and particle concentrations, with quantitative differences evident upon changing the exponent $\alpha$ (governing cluster diffusivity) from $1/2$ to 1. We assume constant coordination number $z_n=2.5$.}
\end{figure}

We next estimate the timescale upon which two clusters of size $n \leq \en $ collide. The cluster diffusion constant is 
\beq
D(n) = D_0 n\, p_{\rm whole}(n),
\eeq
where $D_0$ is the diffusion constant of a monomer. The timescale upon which two clusters of size $n$ encounter each other through diffusion is approximately
\beq
\tau_{\rm enc}(n,n)=\frac{1}{4} D(n)^{-1} \ell_{\rm eff}(n)^2,
\eeq
where
\beq
\ell_{\rm eff}(n) = \left(\phi_0^{-1/d}-2a\right) n^{\gamma}
\eeq
is a measure of the distance separating clusters of size $n$. Here $a=1/2$ is the monomer radius, and $\gamma$ is an exponent measuring the increase in distance between structures due to clustering (our simulations indicate that $\gamma \approx 0.5$ in $d=2$). For brevity we write $\ell_0 \equiv \phi_0^{-1/d}-2a$. Then
\beq
\label{tau_enc}
\tau_{\rm enc}(n,n) = \frac{1}{4} D_0^{-1} n^{\alpha + 2 \gamma} \ell_0^2 \left(1-e^{-\beta \Delta \epsilon} \right)^{1-n}.
\eeq
We compare the cluster-cluster encounter timescale with the timescale $\tau_{\rm evap}(n)$ required for monomers to unbind from clusters of size $n$ and encounter other structures of size $n$:
\bea
\label{tau_evap}
\tau_{\rm evap}(n) &=& \tau_{\rm unbind}(n) + \tau_{\rm enc}(1,n) \nonumber \\
&\approx&D_0^{-1} (n {\cal G}_n)^{-1} e^{\beta z_n \Delta \epsilon}+D_0^{-1} \ell_0^2 n^{2 \gamma}. 
\eea
At low temperature the unbinding timescale is much larger than the timescale for diffusion of a single particle between clusters.
\begin{figure}[h]
\includegraphics[width= 7.5cm]{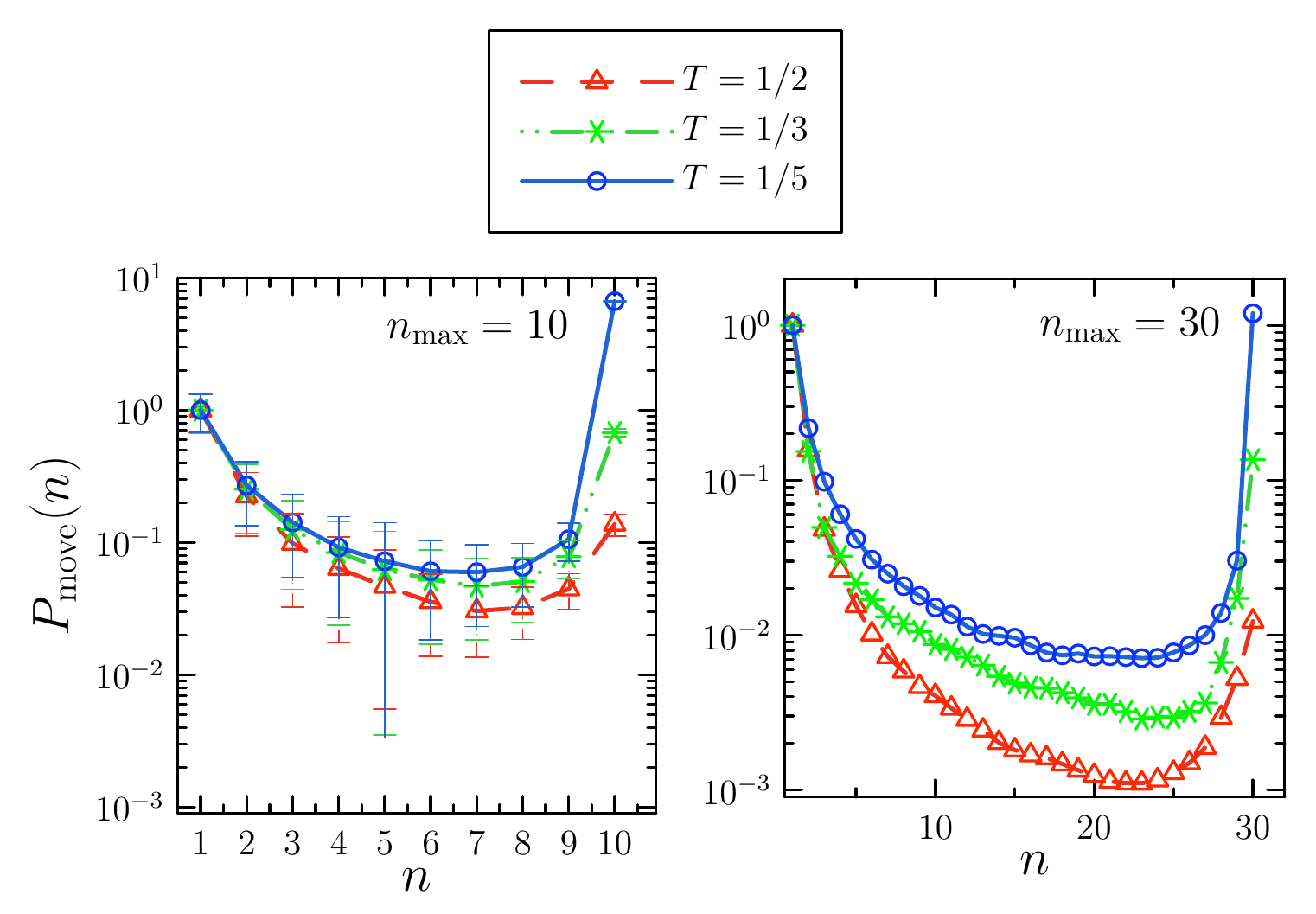} 
\includegraphics[width= 7.5cm]{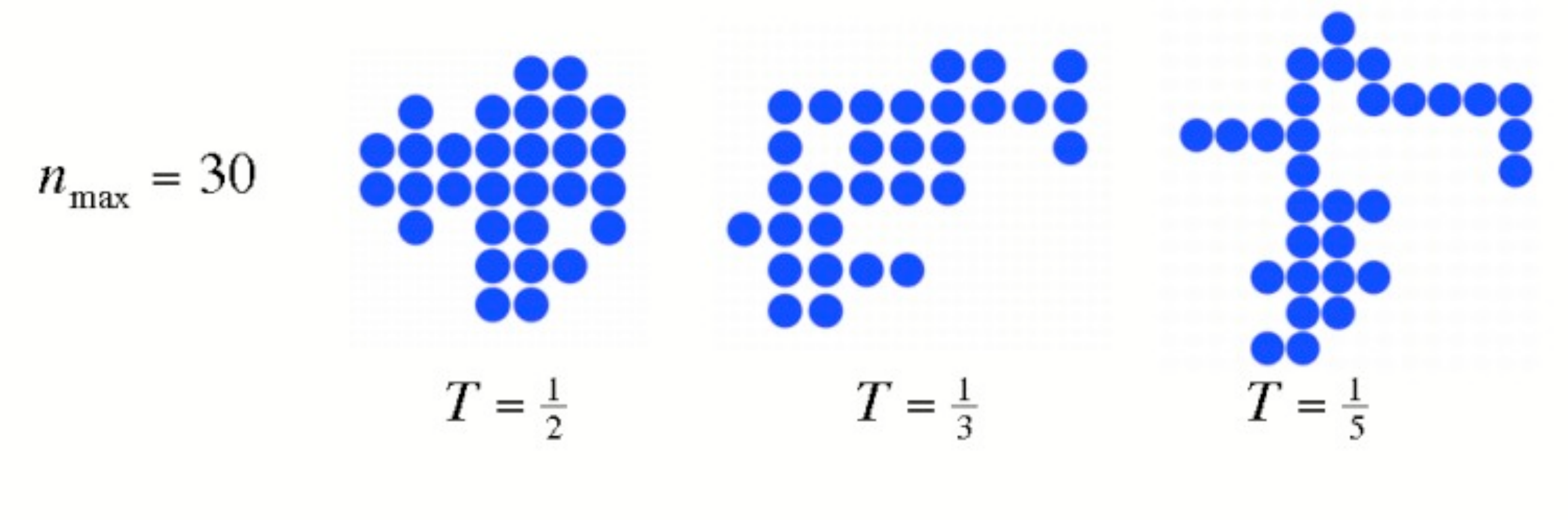} 
\caption{\label{individu} Top row: Kinetic behavior of individual clusters within the $2d$ attractive lattice gas at different $T$. Starting from disordered configurations at concentration $\phi_0=0.1$ we evolve each system according to virtual-move Monte Carlo dynamics with $D(n) \propto n^{-1}$. We `capture' clusters of size $n_{\rm max}=10$ or 30 as they develop, and subject these clusters repeatedly to the same collective-move Monte Carlo procedure, recording (but not making) accepted moves. We plot the probability of the concerted motion of a sub-cluster of size $1 \leq n \leq n_{\rm max}$, $P_{\rm move}(n)$, normalizing data by setting $P_{\rm move}(1)=1$. At the highest temperature the probability of  monomer unbinding, $P_{\rm move}(1)$, is rapid relative to that of whole-cluster motion, $P_{\rm move}(n_{\rm max})$, allowing assembled structures to relax as they grow. Note however that correlated motions contribute measurably to relaxation dynamics even at high temperature. At lower temperatures, collective motion of the whole cluster predominates. Bottom row: typical 30-member clusters obtained at the three temperatures considered.}
\end{figure}

For a cluster of size $n$, Equations~(\ref{tau_enc}) and~(\ref{tau_evap}) quantify the respective timescales for mass transport by collective motion, and for the unbinding and diffusion of monomers. We view the ratio ${\cal P}_{\rm coll}(n) \equiv \tau_{\rm evap}(n)/\tau_{\rm enc}(n)$ as a measure of the propensity for collective motion at a given scale $n$.  The self-assembly of the homogeneous phase proceeds in stages via the appearance of structures of size $n$. At each stage, we expect collective motion to be important if ${\cal P}_{\rm coll}(n)$ is greater than unity. At high temperature, single-particle motion dominates: ${\cal P}_{\rm coll}(n) \propto (\beta \Delta \epsilon)^{n-1} \to 0$ when $\beta \Delta \epsilon \to 0$. At low temperature, instead, collective motion dominates assembly up to large values of $n$: ${\cal P}_{\rm coll}(n) \sim n^{-1+1/d-\alpha-2 \gamma} e^{\beta z_n \Delta \epsilon} \ell_0^{-2}$. For the choices $\gamma=1/2$ and $d=2$, we expect collective motion to be a more effective means of mass transport than single-particle unbinding for clusters of size less than
\beq
n_{\rm max}(\alpha) \sim \left[\ell_0^{-2} \exp \left(\beta \langle z_n \rangle \Delta \epsilon \right) \right]^{2/(2 \alpha +3)},
\eeq
which can be very large at low temperature and low to moderate densities. Under these conditions, whole-cluster motion will dominate the system's assembly dynamics on large length and timescales.
\begin{figure}[h]
\includegraphics[width=0.9\linewidth]{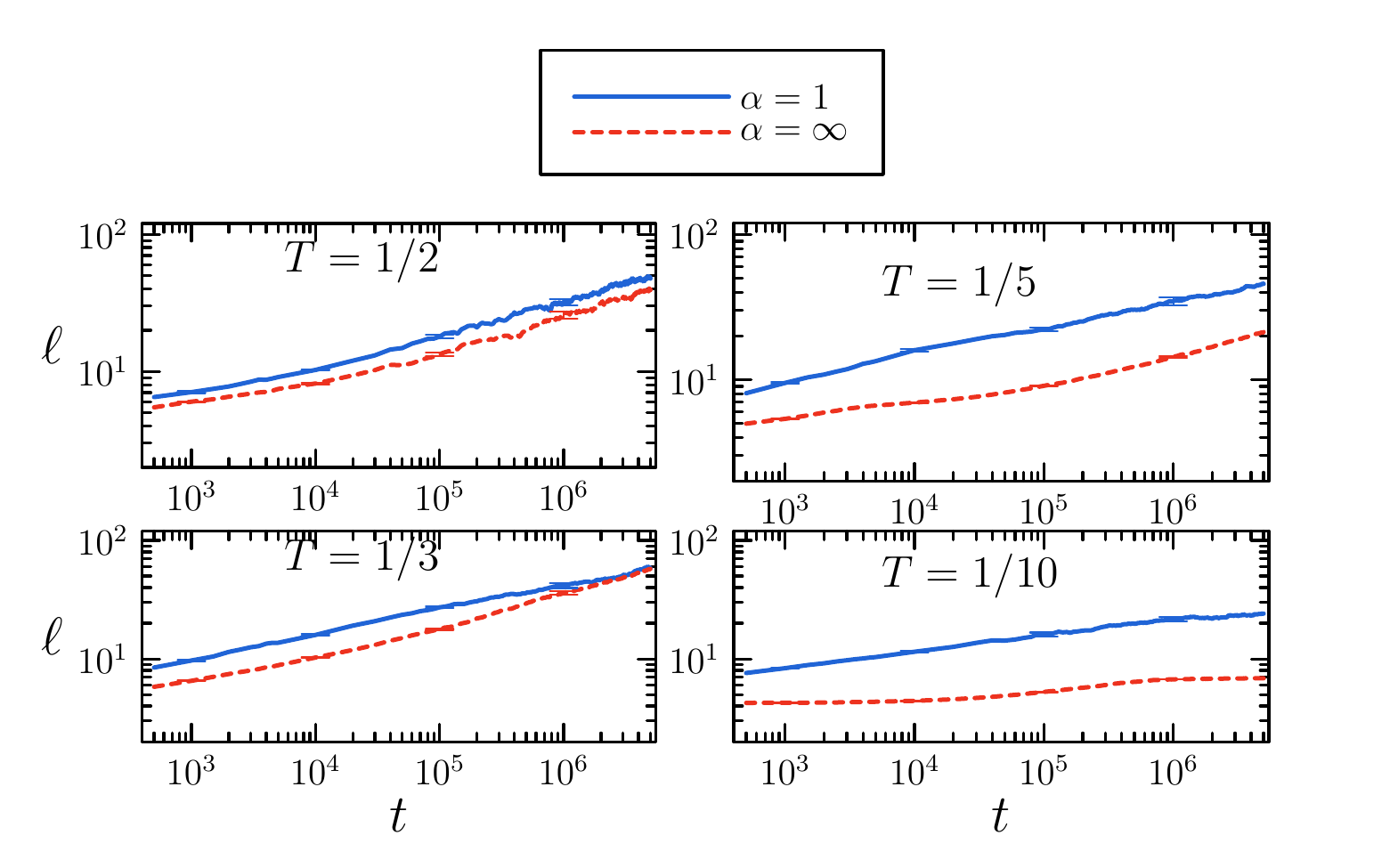} 
\caption{\label{growth1} Growth of domain size $\ell$ with Monte Carlo time $t$ within the attractive lattice gas, for cluster diffusivity $\alpha=1$ or $\alpha=\infty$ at particle concentration $\phi_0=0.1$ on a lattice of size $V=256^2$. We perform quenches to $T=1/2$, $1/3$, $1/5$ and $1/10$. Quantitative differences between dynamical protocols are seen even at the highest temperature. Data correspond to the mean of 10 trajectories; error bars are shown sparsely for clarity.}
\end{figure}

The regimes of single-particle-dominated motion and cluster-dominated motion are separated by the locus of points $(\phi_0^{\star},T^{\star})$, defined for a given $n$ by the equation ${\cal P}_{\rm coll}(n)=1$. We call this locus the `clustodal', by analogy with `binodal' (locus of phase equilibrium), and `spinodal' (locus of the onset of spontaneous decomposition). 

In Figure~\ref{figclus2} we show a plot of the propensity for collective motion, ${\cal P}_{\rm coll}(n)$, in the $(\phi_0,T)$ plane, based on our simple argument. We consider two values of $\alpha$ (quantifying the cluster diffusion rate). The intersections of the clustodal with surfaces of constant $n$ indicate where in state space collective motion dominates assembly dynamics on a scale $n$. Thus for a thermodynamic state $(\phi_0=0.1,T=0.2)$, say, our estimate suggests that assembly dynamics (for $\alpha=1$) for clusters of size 30 (but not size 300) is dominated by motion correlated on times less than the discrete time step $\Delta t$. 

In simulations of the lattice gas we indeed find that collective motion is important for a broad range of temperatures and particle concentrations.  In Figure~\ref{individu} we offer a microscopic perspective on the meaning of the clustodal by analyzing the (averaged) motion of individual lattice gas clusters of sizes $n_{\rm max}= 10$ and 30 at three temperatures. Starting from randomly dispersed monomers at concentration $\phi_0=0.1$, we capture clusters of size $n_{\rm max}$ as they assemble. We use the virtual-move algorithm (enforcing a diffusion constant scaling $D(n) \propto n^{-1}$) to calculate the likelihood of the correlated motion of a subset of size $n$ of a given cluster, normalized by the likelihood of monomer unbinding from that cluster. At high temperature single-particle unbinding is rapid on the timescale of whole-cluster motion, while at low temperature the converse is true. However, even at the highest temperature we find that motions correlated on timescales less then $\Delta t$ contribute appreciably to the `spectrum' of dynamic relaxation. Such motion is ignored within single-particle algorithms. The non-monotonicity of $P_{\rm move}(n)$ reflects the fact that moves of sub-clusters of size $n \approx n_{\rm max}/2$ involve the breaking of many bonds, and so are amongst the least favorable processes in an energetic sense; the asymmetry of $P_{\rm move}(n)$ results from the fact that cluster diffusion constants decrease with increasing $n$. Each data set in Figure~\ref{individu} was obtained by applying $10^7$ moves to a cluster of size $n_{\rm max}$ (recording any accepted move but not making that move), and averaging over $10^3$ such clusters. 

In Figure~\ref{latticepic} we show lattice gas configurations at fixed time following a quench from high temperature to a low final temperature, for four values of this final temperature and for two particle concentrations (here and in subsequent sections we used VMD~\cite{vmd} to render simulation configurations). We employ both collective dynamics $\alpha=1$ and single-particle dynamics $\alpha=\infty$. The equilibrium states (`equil.') are in all cases phase-separated configurations, which we deduce by using nonlocal Monte Carlo moves~\cite{chen}. At the highest temperatures, coarsening dynamics are visually similar. In this regime the system lies `outside' the clustodal for all but the smallest values of $n$, and single-particle evaporation and diffusion is the dominant mode of relaxation. At lower temperatures and the lower of the two particle concentrations, striking differences emerge: for sufficiently low temperatures collective modes of motion ($\alpha=1$) induce kinetic frustration through the mutual collisions of clusters that bind awkwardly and fail to relax before encountering similar structures. If collective motion is suppressed $(\alpha=\infty)$ then instead isolated, compact structures are formed. At the higher particle concentration, steric effects partially frustrate large-scale collective motion, and coarsening patterns are qualitatively more similar (although differences can be clearly seen).

We quantify the influence of collective motion upon assembly dynamics by measuring the characteristic lengthscale of domains as a function of time at $\phi _{0} = 0.1$, for $\alpha=\infty$  and $\alpha=1$. We calculate the domain lengthscale $\ell $ from the first moment of the structure factor~\cite{amar}; we display in Figure~\ref{growth1} results for four temperatures.  At the highest temperature, $T=1/2$, single-particle and cluster algorithms show the same qualitative behavior, with both $\ell $ and $d\ell /dt $ increasing monotonically with time. Both dynamics produce compact clusters, as may be seen in Figure~\ref{latticepic}. However, even at this high temperature we observe quantitative differences between the two algorithms, with collective motion giving rise to larger domains at any given time than does single-particle motion. The mean cluster size does not become large enough that whole-cluster diffusion is negated as a viable means of assembly. Interestingly, and counter-intuitively, the growth in lenghscale under the two algorithms is more similar at late times at a slightly {\em lower} temperature of $T=1/3$. Here we observe the same qualitative assembly behavior under the two algorithms as at $T=1/2$, but now the mean cluster size is larger at late times because of the stronger thermodynamic driving force. Even when collective dynamics is permitted, collisions involving these larger clusters are sufficiently rare that increases in $\ell$ are driven chiefly by single-particle unbinding events; the two algorithms behave similarly in this regime.

At the two lowest temperatures, $T=1/5$ and $1/10$, single-particle dynamics displays the 
same qualitative behavior as it does at $T=1/2$ and $1/3$, but cluster moves show different behavior.  While $\ell $ still increases monotonically with time, the slope $d\ell /dt $ has at late times a constant value at $T=1/5$, and decreases at $T=1/10$. The corresponding pictures for $\phi _0 = 0.1 $ in Figure~\ref{latticepic} suggest that this decrease in slope results from the steric hinderance associated with percolating fractal-like clusters. Such clusters coarsen chiefly by way of rare single-particle unbinding events, a mechanism that leads to smaller increases in domain length with time than do mutual cluster-cluster collisions.

Our results show that the $2d$ attractive lattice gas displays a rich variety of coarsening or self-assembly behaviors, and that these behaviors depend sensitively upon the dynamical protocol used to evolve the system. Figure~\ref{latticepic} demonstrates that when one accounts for collective motion $(\alpha=1)$, nucleation and growth mechanisms ($T=1/2,\phi_0=0.1$) yield to gelation as temperature is reduced ($T=1/5$ and $1/10,\phi_0=0.1$). Gelation at low temperature is driven by explicit collective motion, or motion correlated on the fundamental timescale $\Delta t$: when we forbid such motion $(\alpha=\infty)$, gelation is avoided in favor of very slow nucleation and growth. It is interesting that collective motion  appears to be crucial to the formation of large-lengthscale gels within the attractive lattice gas at low particle concentration, but that such explicit correlated motion is not required to observe glassy behavior within a model of (dense) silica~\cite{mc_dynamics4}. These results imply a difference in the nature of the dynamical cooperativity associated with glasses and attractive gels, respectively, short-time cooperativity versus long-time cooperativity.

\section{The attractive lattice gas generalized to continuous space: model associating colloids}
\label{sec_colloid}

The effects of collective motion seen in the lattice gas are also apparent in a simple off-lattice model of hard spheres equipped with isotropic interactions. Such a model is a caricature of colloidal particles in solution with small polymers; the colloids associate by virtue of a polymer-mediated depletion attraction. In Ref.~\cite{lu}, attractive colloidal particles of this nature were observed in experiment to assemble into clusters whose geometry depended upon the range and strength of the depletion attraction. Strikingly, at low colloid concentration these clusters appeared to be stable, defying the expectation that components with strong pairwise attractions should phase-separate or gelate. 

We consider as a simple model of this system a three-dimensional collection of $N$ hard spheres of diameter $\sigma$ equipped with an attractive square well of range $\xi \sigma$ and strength $U$. Spheres are placed randomly within the simulation box and occupy 4\% of its volume. We impose periodic boundary conditions. We consider two of the parameter sets described in Ref.~\cite{lu}: a potential of moderate strength and range, A $(U=2.6 \, \kt, \xi = 0.11$), and a potential of considerable strength and short range, B $(U=12\, \kt, \xi = 0.02$). We carried out Monte Carlo simulations of systems of $N=1085$ and 6500 particles, drawing particle displacement magnitudes (in units of $\sigma $) uniformly from the interval $[0,0.15] $, and rotation angles uniformly from a distribution with maximum $\sim 14^{\circ}$. We used `virtual-move' translations (employing the algorithm described in the appendix), and used a static linking scheme to effect rotations about the center of mass of a chosen pseudocluster. These choices of displacement magnitude and rotation angle imply a basic timescale $\Delta t$ sufficiently large that to a good approximation large-scale cooperative motion cannot occur from uncorrelated moves of single particles; such moves effect only local structural relaxation and the binding, unbinding and diffusion of monomers. Our aim is to compare such dynamics with pathways accessible to explicitly correlated motion. When we consider cluster moves we also compare `freely-draining' with `Stokesian' cluster diffusion scalings. Freely-draining motion implies $(D_{\rm trans.} \propto n^{-1}, D_{\rm rot.} \propto I^{-1})$, where $n$ is the number of monomers comprising the cluster and $I$ is the cluster moment of inertia about the rotation axis. For Stokesian scalings we take $(D_{\rm trans.} \propto R^{-1}, D_{\rm rot.} \propto R^{-3})$, with $R$ a measure of the cluster radius of gyration perpendicular to the translation vector or axis of rotation.

\begin{figure}[h]
\includegraphics[width= 6cm]{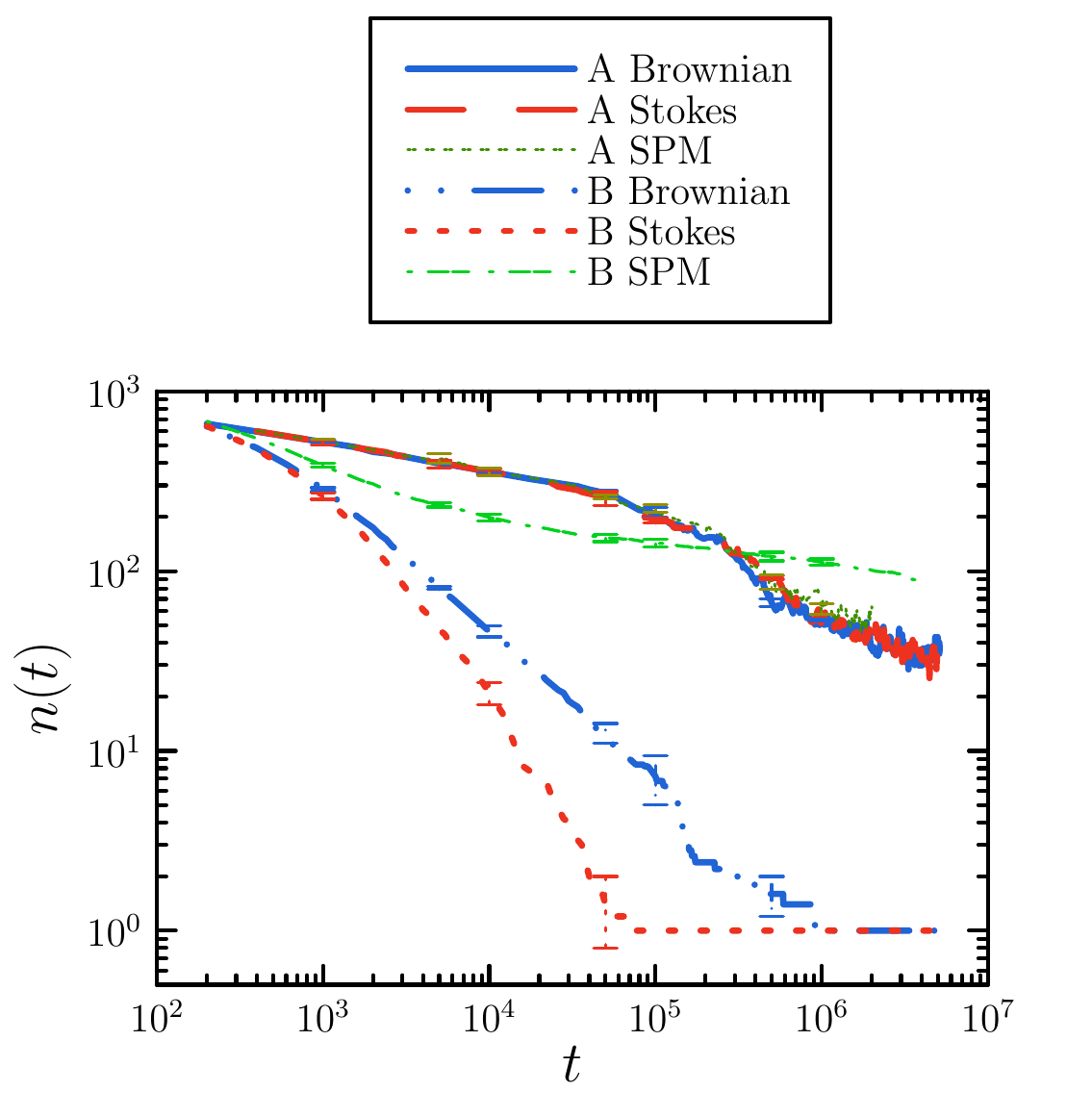} 
\caption{\label{growth} Kinetic data for the self-assembly of model colloids. We show number of clusters $n$ as a function of Monte Carlo time $t$ for systems A $(U=2.6 \, \kt, \xi = 0.11$) and B $(U=12\, \kt, \xi = 0.02$), under `freely-draining' (Brownian) and `Stokesian' cluster diffusion scalings. System A assembles or coarsens chiefly through binding, unbinding and diffusion of monomers, and displays little dependence upon cluster diffusivity. These data superpose on data generated using a single-particle Monte Carlo algorithm (SPM). System B instead assembles through the concerted motion of aggregates (when collective motion is permitted), and displays a dependence upon cluster diffusivity. When denied collective motion, clusters coarsen via very slow monomer unbinding events. Data are averaged over 5 stochastic trajectories, each generated using 1058 particles. Error bars are displayed sparsely for clarity.}
\end{figure}

\begin{figure}[h]
\includegraphics[width= 6cm]{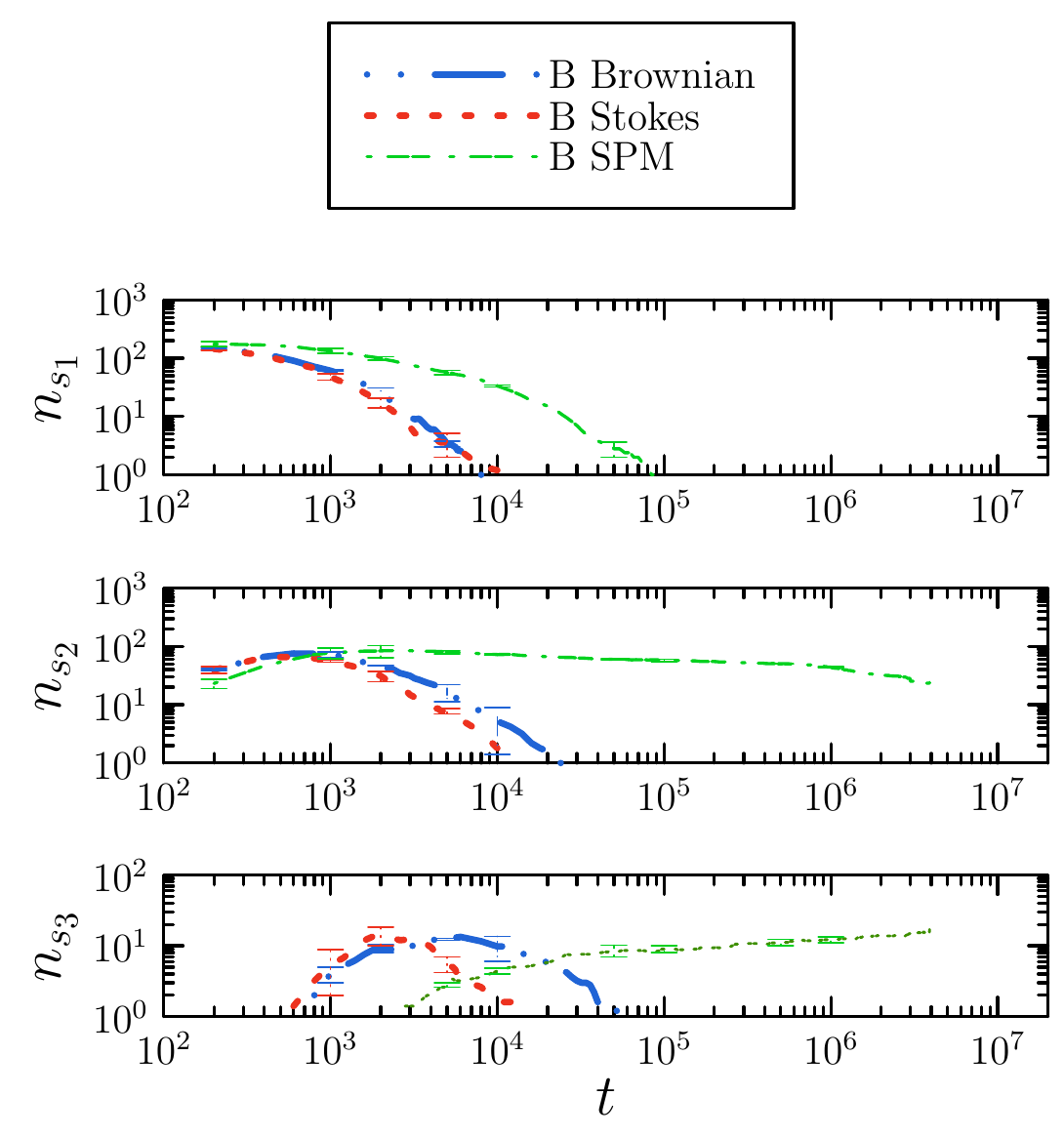} 
\caption{\label{growth2} Kinetic data for the self-assembly of model colloids. We show the number of clusters of given set of sizes $s_i$, $n_{s_i}$, as a function of Monte Carlo time $t$ for system B under Brownian and Stokesian cluster diffusion scalings, and for single-particle moves (SPM). Here $s_1=3$, $s_2$ denotes cluster sizes from 7 to 10, and $s_3$ denotes cluster sizes from 26 to 64. Dynamical pathways for the two realizations of collective motion differ even for relatively small aggregates. Data are averaged over 5 stochastic trajectories, each generated using 1058 particles. Error bars are displayed sparsely for clarity.}
\end{figure}
 
As in the lattice gas, collective motion plays a dominant role when potential energy gradients encountered on a timescale $\Delta t$ are large, and is less important than single-particle unbinding events for small energy gradients. In Figures~\ref{growth} and~\ref{growth2} we show kinetic measures for systems of 1058 particles. System A displays no statistically significant difference in the evolution of the number of clusters with time if collective motion is allowed or suppressed. System B, however, experiences dramatically different fates under collective and individual particle motion, with the former inducing gelation and the latter resulting in a slowly-coarsening collection of isolated clusters. 

\begin{figure}[ht]
\includegraphics[width= 4cm]{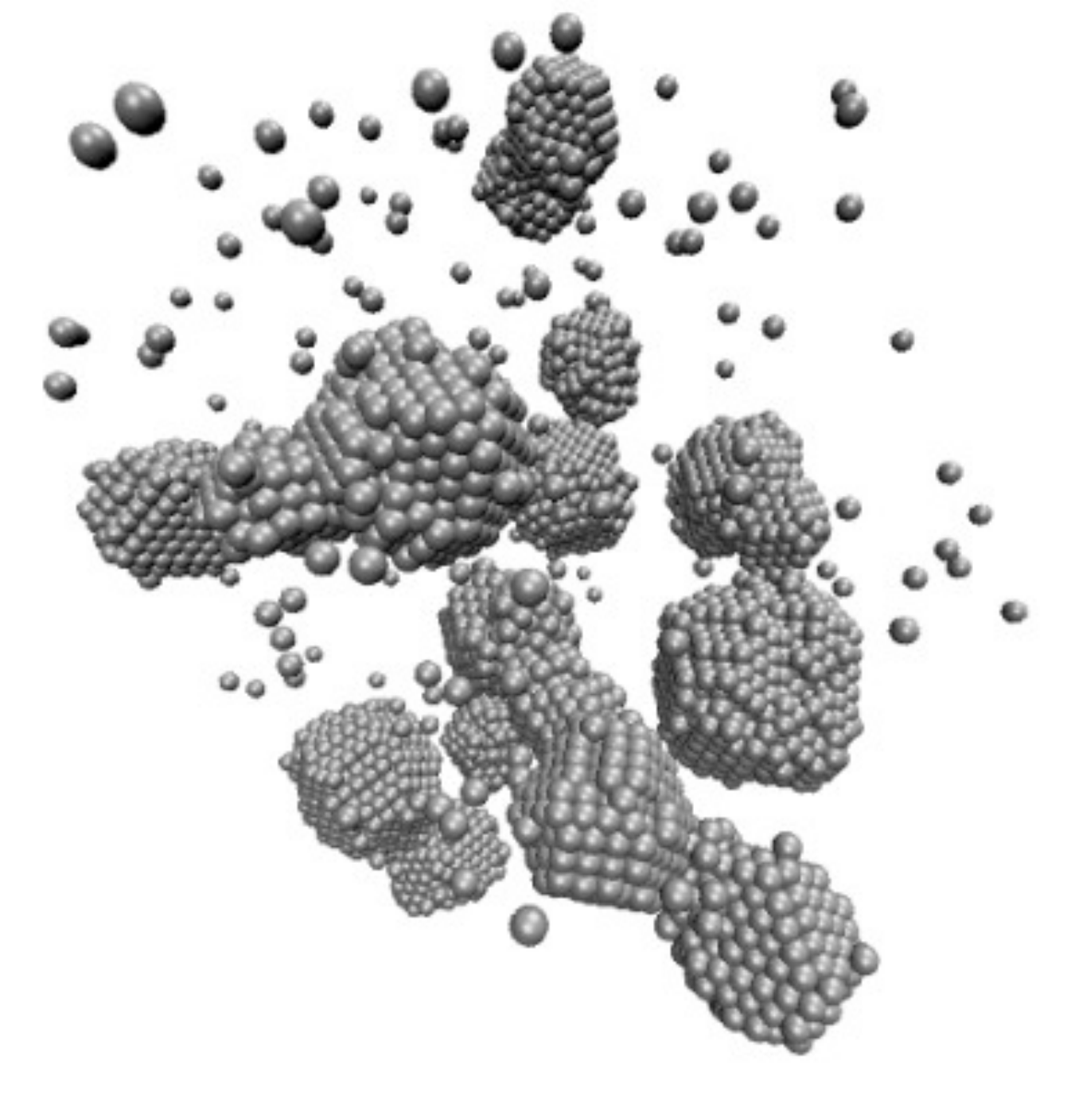} 
\includegraphics[width= 4cm]{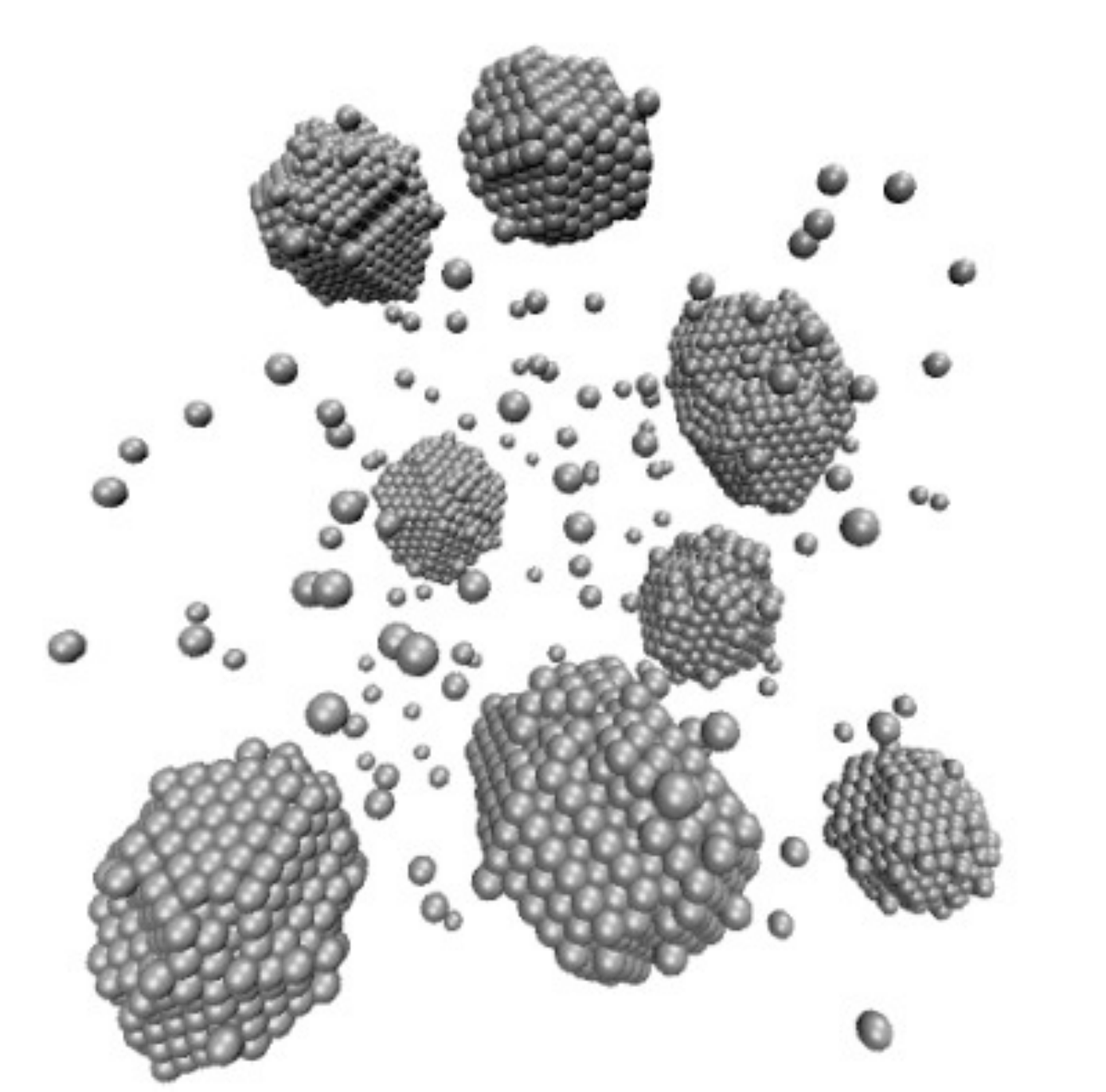} 
\includegraphics[width= 4.5cm,angle=90]{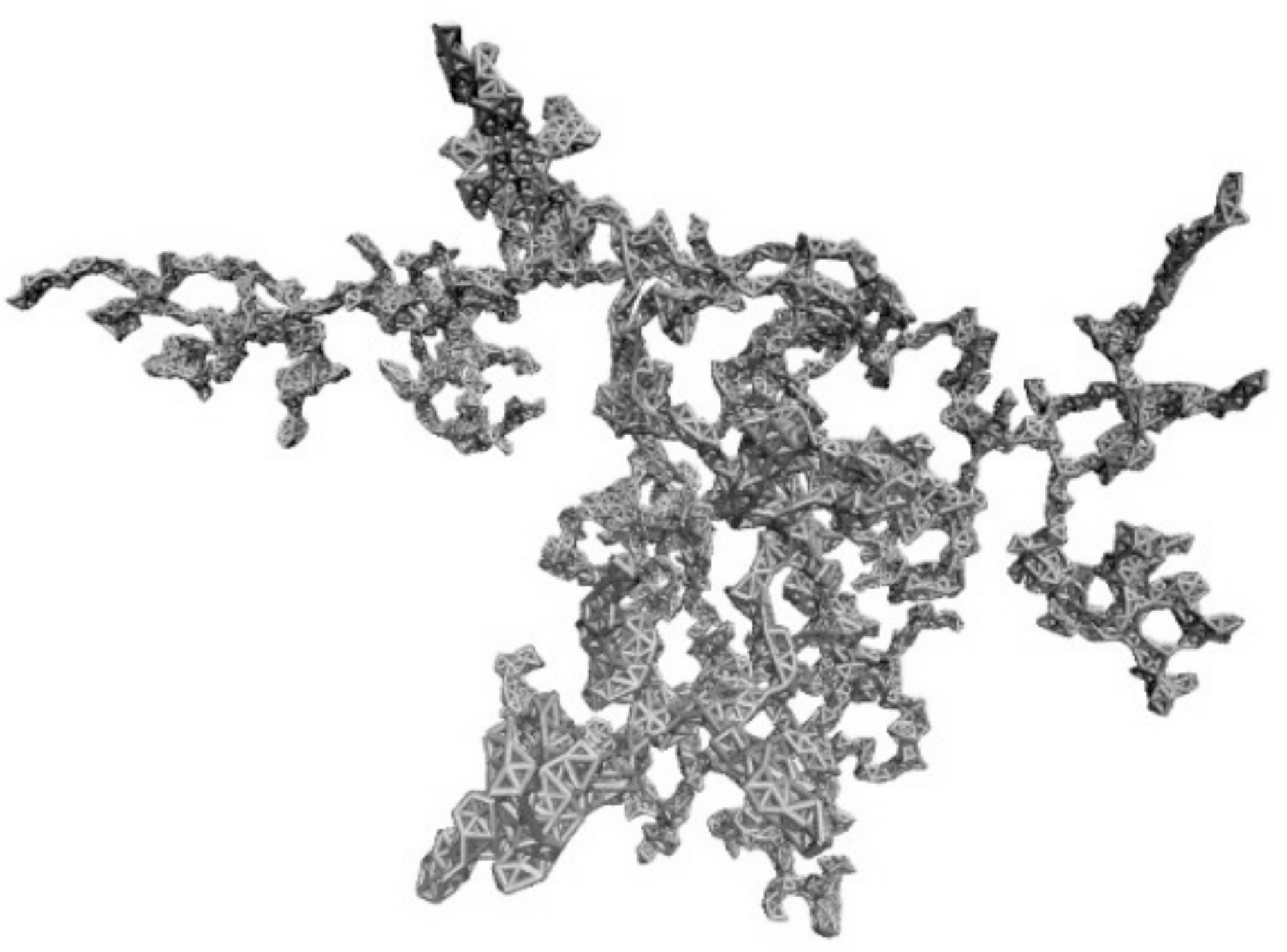} 
\includegraphics[width= 3.7cm]{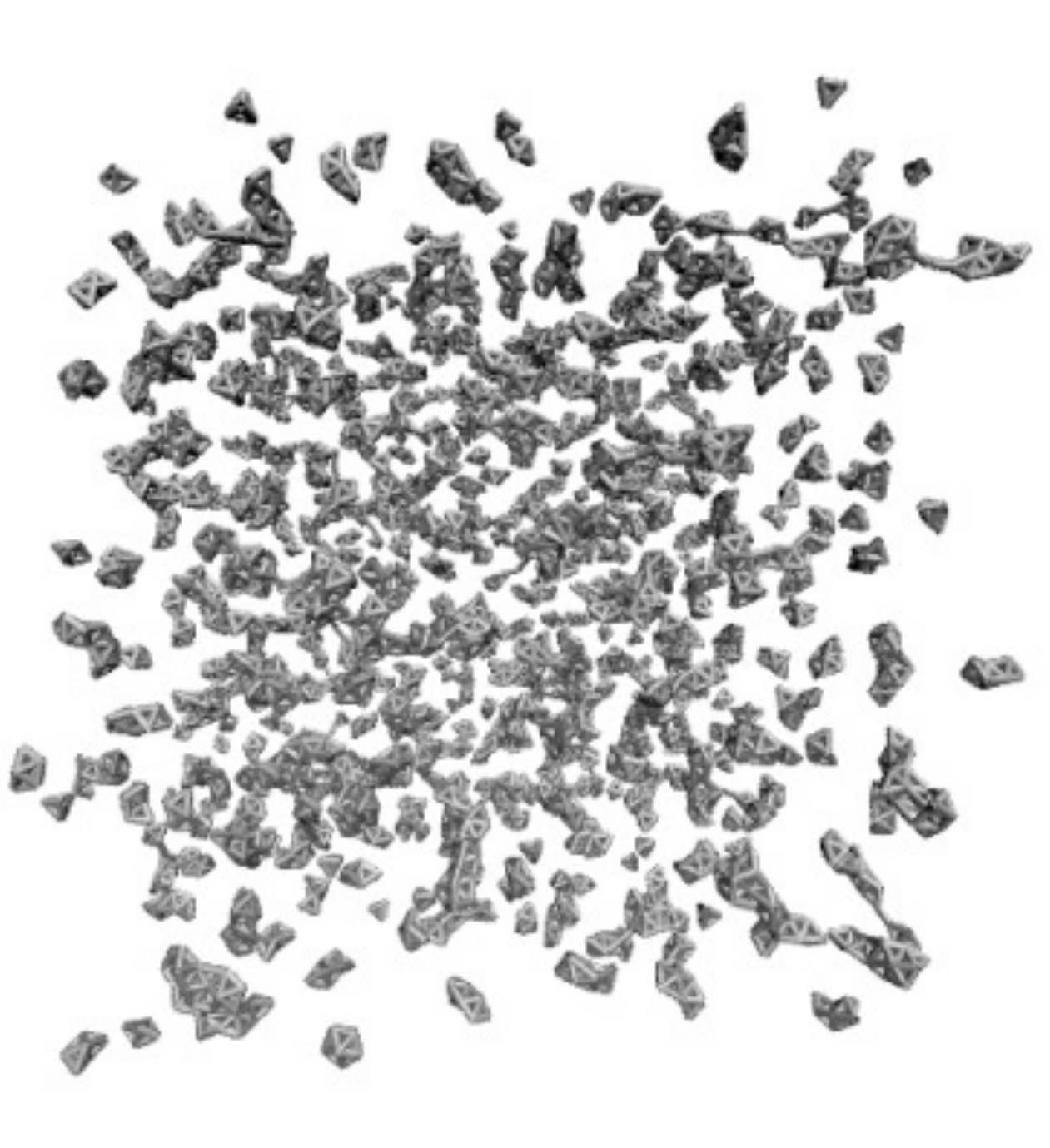} 
\caption{\label{figcoll} Structures obtained from Monte Carlo simulation of 6500 particles equipped with attractive interactions of type A (top row, excluded-volume view) and B (bottom row, bond view), permitting (left column) or forbidding (right column) explicit collective  (Stokesian) motion. The behavior of system A differs under the two dynamical protocols only at late times, where collective motion leads to the aggregation of crystalline clusters. For system B, collective motion induces gelation while single-particle motion results in small, slowly-ripening clusters. These behaviors are similar to those of the lattice gas (see Figure~\ref{latticepic}). Times of image capture (clockwise from top left), in units of $10^6$ MC steps, are 1.4, 2.9, 1.7 and 0.6.}
\end{figure}

\begin{figure}[ht]
\includegraphics[width= 2.3cm]{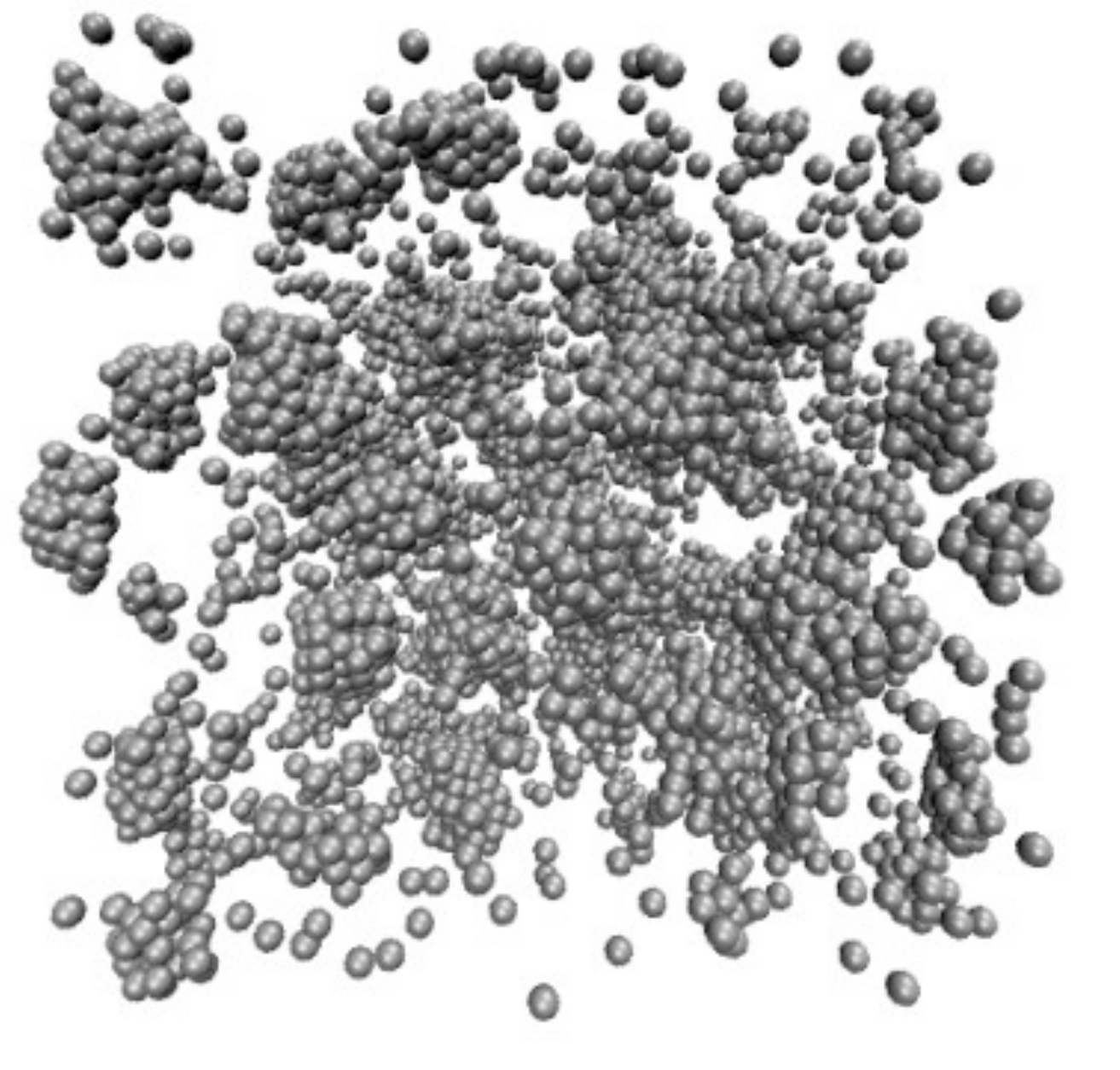} 
\includegraphics[width= 2.3cm]{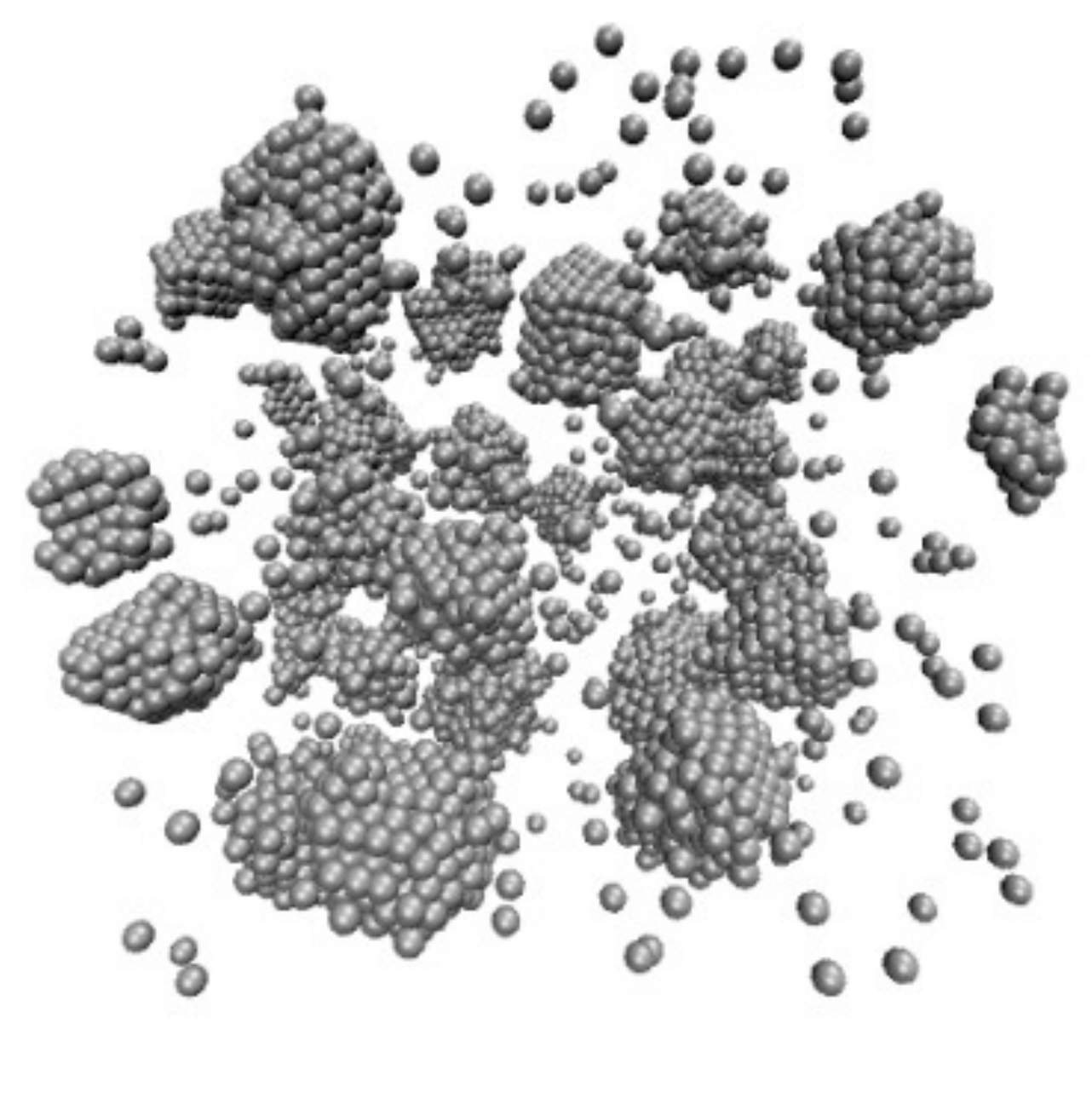} 
\includegraphics[width= 2.3cm]{ep2p6_vmmc} 
\caption{\label{figcoll2} Assembly pathway for system A under explicit collective moves. Assembly is driven principally by single-particle binding, unbinding and diffusion, with the effects of collective motion apparent only at late times when phase-separated clusters fuse. Times of image capture, in units of $10^6$ MC steps, are from left to right 0.2, 0.6 and 1.4.}
\end{figure}

In Figures~\ref{figcoll} and~\ref{figcoll2} we display snapshots of these systems for collections of 6500 particles (in these images system A was evolved using a maximum translation of 0.3$\sigma$). System $A$ undergoes phase separation into crystalline clusters. The effect of collective motion is apparent only at late times when clusters fuse; such fusing is in qualitative agreement with Brownian dynamics simulations of a similar system~\cite{patrick}. System B forms under collective motion stringy, frustrated aggregates that merge and form a gel, with the timescale for gelation different for the two different cluster diffusivities modeled. In experiment, colloids with a strong depletion attraction (of nature similar to systems A and B considered here) form instead isolated clusters~\cite{lu}, and do not phase-separate or gel. The authors of Ref.~\cite{patrick} discuss possible reasons for the disparity between this experimental observation and the cluster-cluster aggregation seen in simulations in Ref.~\cite{patrick} (similar to those observed here): one such suggestion is that, in experiment, an accumulation of charge on large bodies might induce a repulsion that stabilizes a phase of isolated clusters.

\section{A model of viral capsid self-assembly} 
\label{sec_capsid}

The ability of interacting proteins to spontaneously form icosahedral capsids {\em in vivo} and under some conditions {\em in vitro} is a striking example of biological self-assembly. Protein subunits assemble under a variety of conditions, avoiding both kinetic and thermodynamic traps. Understanding the mechanisms that render assembly so robust is an essential step towards designing synthetic analogs of viral capsids. In addition, a comprehensive understanding of the viral capsid formation process will spur the design of antiviral drugs and new drug delivery systems; the latter could possess the ability to assemble and disassemble around their cargo without requiring explicit external control. 

In Ref.~\cite{mike} a class of simple models was introduced in order to study the mechanism by which interacting protein subunits might form `capsids', 60-member closed shells having icosahedral symmetry. We focus here on the `B3' model of that reference. Units interact via a pairwise potential that models an excluded volume and a short-ranged, angularly specific attractive interaction designed to stabilize capsids. 

\begin{figure}[h]
\includegraphics[width=8cm]{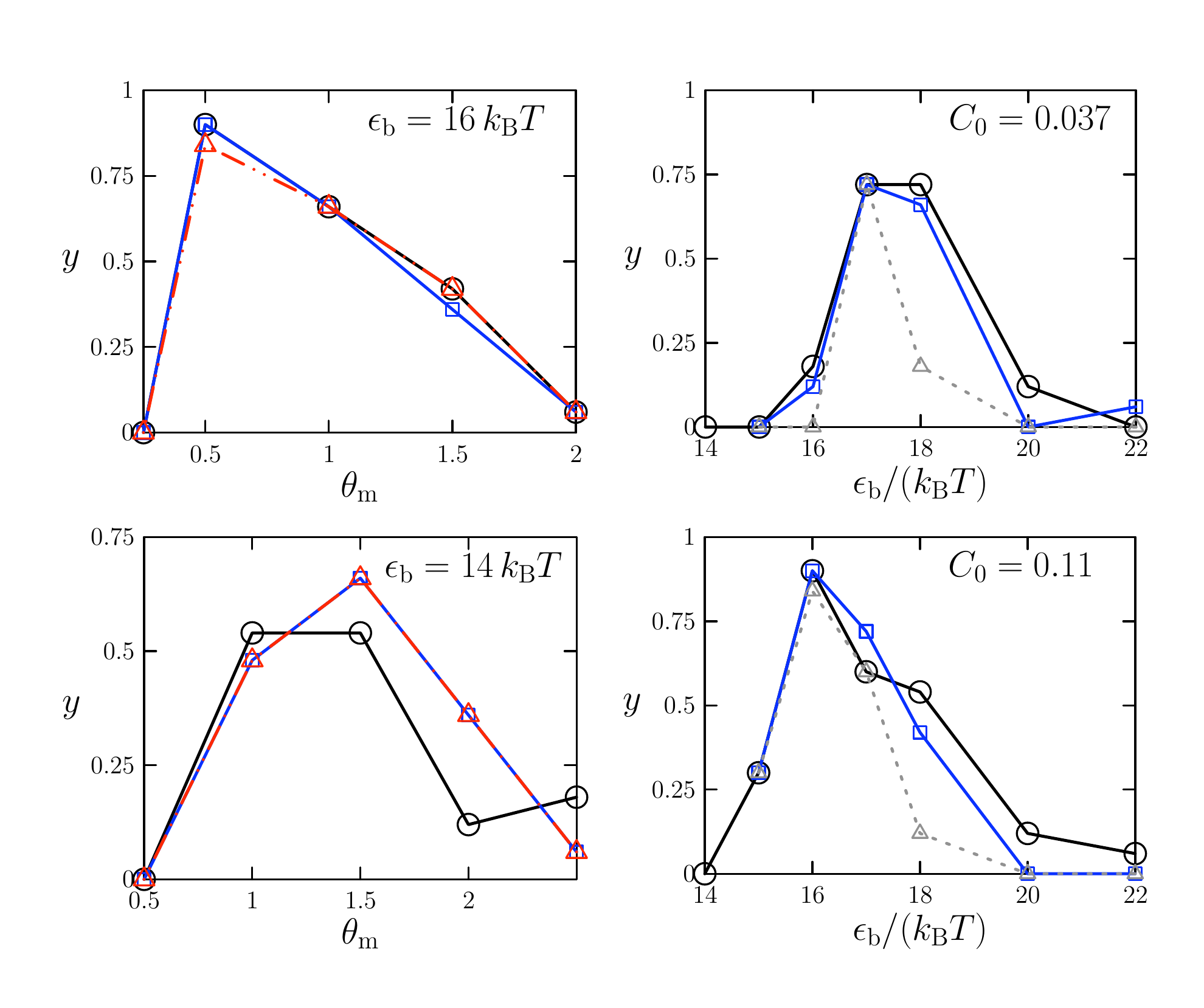} 
\caption{\label{figyield} Final yield $y$ of capsid model as a function of potential specificity $\theta_{\rm m}$ (left column) or strength $\epsilon_{\rm b}$ (right column). Left panels: yield at subunit concentration $C_0\equiv N \sigma^3/L^3=0.11$ for attraction strength 16 $\kb T$ (top left) or 14 $\kb T$ (bottom left); Right panels: yield at binding specificity $\theta_{\rm m}=0.5$ for concentration $C_0=0.037$ (top right) or $C_0=0.11$ (bottom right). Circles denote Brownian dynamics results~\cite{mike}; squares denote results obtained using collective Monte Carlo dynamics; triangles denote results obtained using single-particle Monte Carlo dynamics at times equal to (dashed lines, right column) those at which collective-move data were sampled, or at later times (if available) when yield had appeared to saturate (dot-dashed lines, left column). Times of data capture for Monte Carlo simulations: left panels, 4.2$\times 10^6$ MC steps; right panels 9$\times 10^6$ MC steps. Exceptions are the late-time single-particle data captured at 9$\times 10^6$ MC steps (top left panel, triangles) and $7.1\times 10^6$ MC steps (bottom left panel, triangles).}
\end{figure}

\begin{figure}[h]
$\begin{array}{cc}
\includegraphics[width= 4cm]{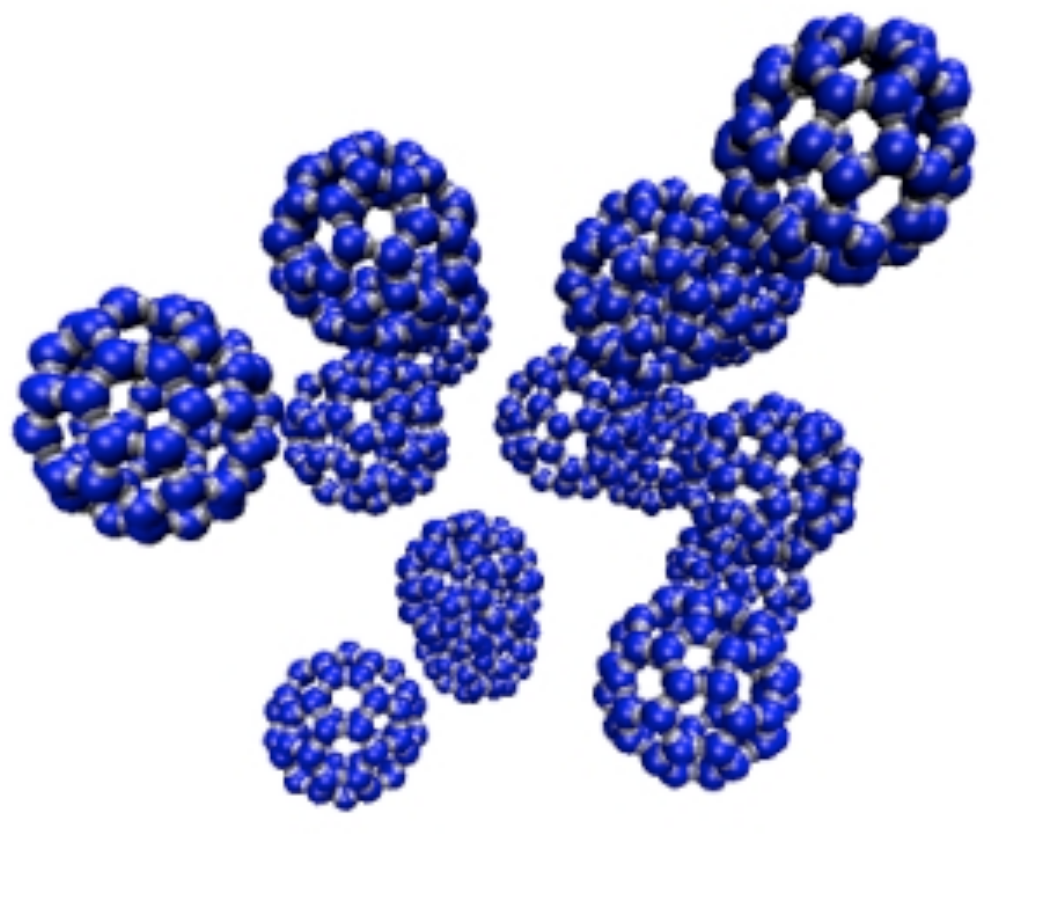} &
\includegraphics[width= 3cm]{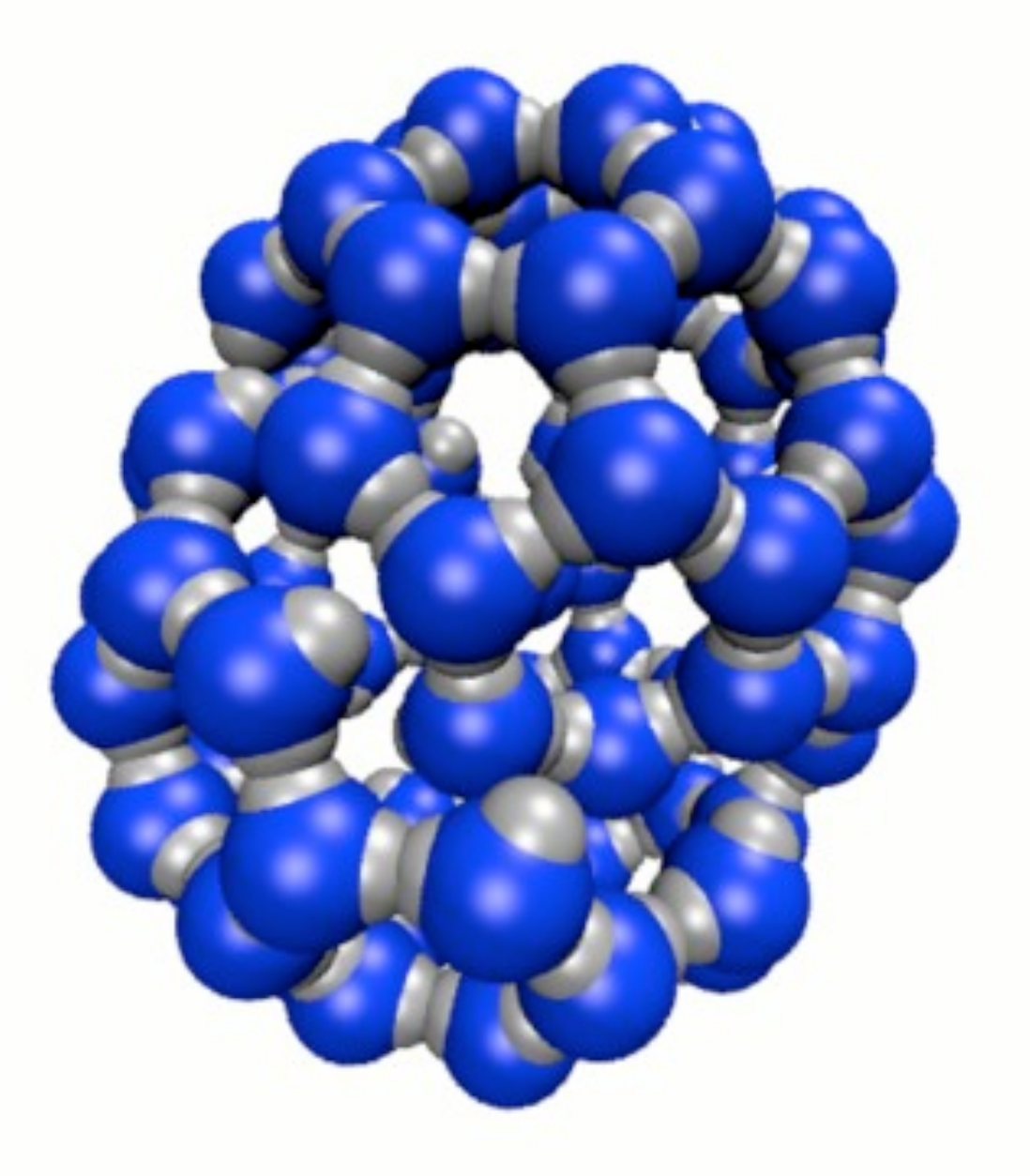} 
\end{array}$
\caption{\label{figsnap} Configurations of the capsid model generated using virtual-move Monte Carlo. Left: configuration illustrating high yield obtained at parameter set $\epsilon_{\rm b} = 16 \, \kt$, $\theta_{\rm m} = 0.5$, $C_0=0.11$. Right: malformed shell of 76 particles obtained at parameter set $\epsilon_{\rm b} = 22\, \kt$, $\theta_{\rm m} = 0.5$, $C_0=0.037$.}
\end{figure}

Here we examine the role of collective motion within the assembly dynamics of this model. We evolved via virtual-move Monte Carlo a collection of 1000 subunits. These are initially randomly oriented and dispersed within a three-dimensional simulation box; we take periodic boundaries in each dimension. We scaled cluster diffusivities in order to approximate Brownian dynamics. As found in Reference~\cite{mike}, we observe a regime of inter-unit potential strength and specificity within which assembly is robust. We show in Figure~\ref{figyield} the capsid yield (fraction of units residing in complete capsids) at fixed observation time as a function of potential strength $\epsilon_{\rm b}$ or specificity $\theta_{\rm m}$. Yield data obtained via virtual-move Monte Carlo agree with Brownian dynamics results to within statistical error (considerable variations in yield are observed within each algorithm at a given thermodynamic state). Particle displacement magnitudes are drawn from a uniform distribution with maximum equal to a length unit $\sigma$; particle interaction range is $2.5 \sigma$. Rotations are scaled accordingly. Yields are non-monotonic functions of these parameters, for the reasons outlined in Figure~\ref{figcomp}: overly strong or insufficiently specific interactions promote malformed intermediates that fail to assemble into complete structures; overly weak or specific interactions result in productive subunit-subunit binding events that are too rare to induce assembly on the timescales simulated. We show in Figure~\ref{figsnap} example configurations obtained from a well-assembled and a badly-assembled system.

\begin{figure}[h]
$\begin{array}{cc}
\includegraphics[width= 8cm]{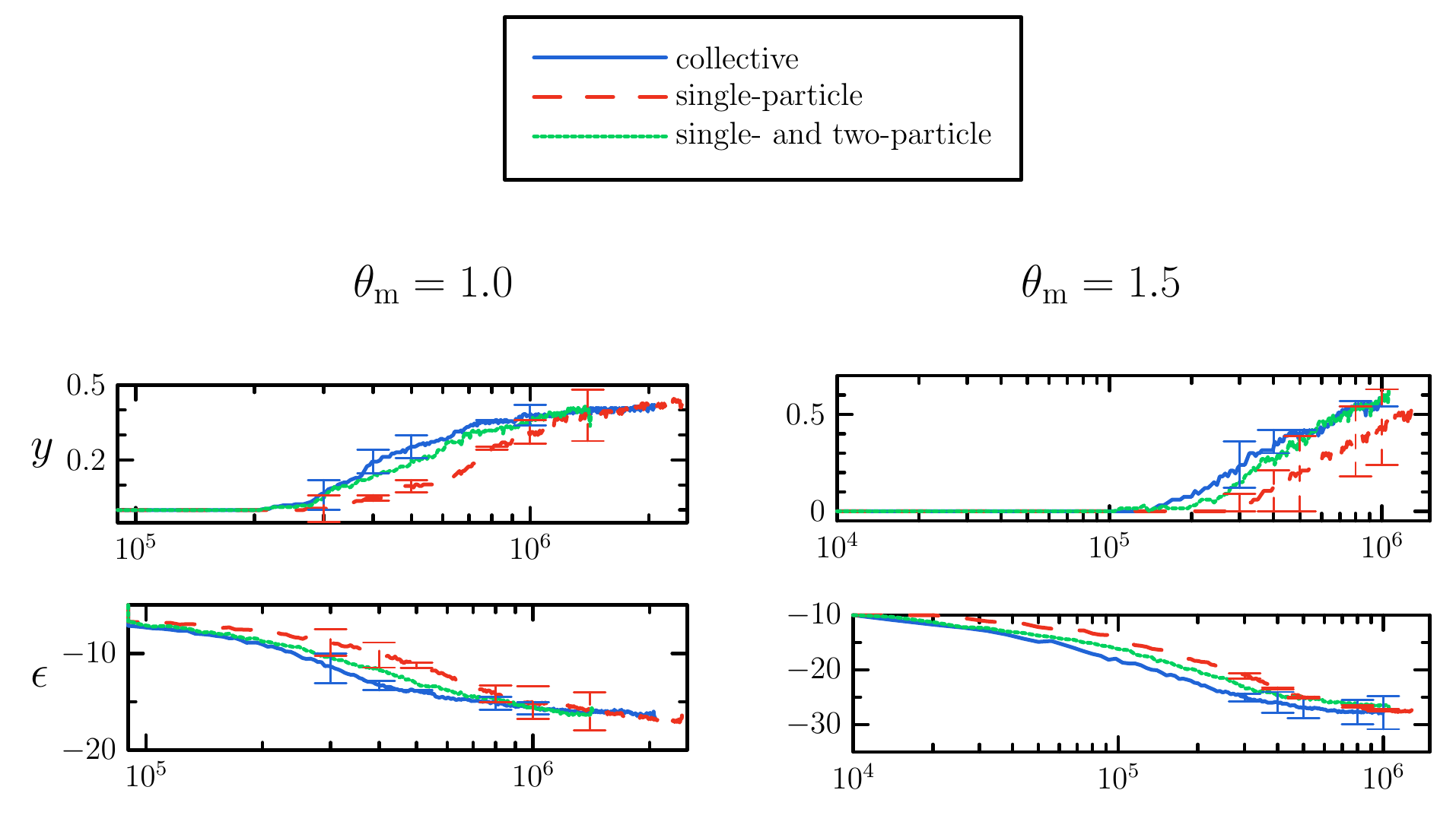} &
\end{array}$
\caption{\label{figcompare} Capsid kinetics at binding energy $\epsilon_{\rm b} = 14 \, k_{\rm B} T$ and concentration $C_0=0.11$ evolved using Monte Carlo dynamics with different degrees of collective motion permitted. We show yield $y$ and energy per particle $\epsilon$ as a function of time. Assembly under collective moves is more efficient than under single-particle moves. Restoring collective motion of dimers recovers a substantial fraction of the efficiency of fully collective motion. Data in left panel are the mean of 10 trajectories, data in right panel are the mean of 4 trajectories. Error bars are displayed sparsely for clarity.}
\end{figure}

We can examine the role of collective motion in this example of self-assembly by explicitly restricting or forbidding collective modes of relaxation within Monte Carlo dynamics. We find that in many cases final yields are not strongly affected by doing so, but that assembly dynamics are impaired. In Figure~\ref{figyield} we show yields obtained using single-particle Monte Carlo dynamics (triangles) at Monte Carlo times equal to (dotted lines) the times at which collective-move Monte Carlo data were sampled, and for times at which yields appeared to saturate (dot-dashed lines). Saturated yields were not obtained (after 300 hours of simulation) for single-particle moves in the right panels of  Figure~\ref{figyield}.  

In Figure~\ref{figcompare} we present measures of assembly kinetics obtained at two parameter sets for fully collective motion (we move clusters according to the algorithm described in Section~\ref{section_algorithm}, with the diffusivity of $n$-mers chosen to be $D(n) \propto n^{-1}$ in order to approximate Brownian dynamics), for motion allowing explicit moves of monomers and dimers only (as for fully collective motion, but with the diffusion constant of trimers and higher-order clusters set equal to zero) and for single-particle motion (diffusion constant of dimers and higher-order structures set equal to zero). To generate these data we drew particle displacement magnitudes from a uniform distribution with maximum equal to $0.2\sigma$. While at these thermodynamic states the final yields do not depend strongly upon the availability of correlated moves, the kinetics of assembly is more rapid when collective motion is allowed. Interestingly, restoring explicit moves of dimers alone is sufficient to recover much of the ease of assembly afforded by a `full' spectum of correlated motions. These results support the observations made in Ref.~\cite{mike}. There it was found that B3 capsids grow in part through events consisting of collisions between intermediates larger than monomers, implying that collective motion plays an important role in the model's assembly dynamics (see also Ref.~\cite{Zhang2006}). Further, the most frequent intermediate binding events for the B3 model involved dimers.  Indeed, we find here that explicit dimer motion renders assembly much more facile than if moves are uncorrelated. We note also that the tendency of subunits to form closed shells suppresses in large part the kinetic trapping seen in the other models studied in this paper, where large aggregates bind in an awkward fashion and frustrate equilibration.

These capsid assembly results may be contrasted with the assembly properties of a model of interacting protein complexes called chaperonins, studied in Ref.~\cite{vmmc}. There the non-complementarity of model chaperonin-chaperonin interactions coupled with collective modes of motion induce a high degree of kinetic frustration in some regions of the phase diagram: large clusters bind awkwardly, which slows or prevents equilibration. Single-particle moves, which suppress such cluster diffusion, give rise to small, isolated, well-formed structures. We conclude that particle interaction geometry and the extent of organized aggregates play a decisive role in shaping the effects of collective motion in self-assembling systems.

\section{Conclusions}

We have examined the role of collective motion in Monte Carlo simulations of three model systems. We find that collective motion is responsible for gelation at low particle concentrations within the two dimensional lattice gas, at thermodynamic states for which the equilibrium configurations are phase-separated. Suppressing collective motion at the same thermodynamic states instead leads to nucleation and growth of clusters. Collective motion plays a similar equilibration-frustrating role within a simple off-lattice model of associating colloids in three dimensions, driving gelation when interactions are strong. By contrast, correlated motions of anisotropically-interacting subunits within a model of viral capsid assembly lead to more efficient self-assembly under all conditions considered. Our results suggest that gelation of homogeneously-interacting particles might be regarded as coarsening in the presence of collective motion, and that the interplay of phase-separation and gelation depends both upon the thermodynamic state, and upon the rate of diffusion of self-assembled aggregates.

Control of collective motion in real systems is very difficult, but might be possible in special circumstances. Individual colloids in high-salt polymer solutions interact via only short-ranged depletion attractions, but buildup of charge on large aggregates can lead to many-body electrostatic repulsions. It is possible that fine-tuning of solution conditions could be used to select the smallest aggregate lengthscale for which repulsions become significant, thereby controlling the extent to which multimers collide and bind. Many-body effects of a different nature can be induced by long-ranged hydrodynamic interactions in, for example, sedimenting colloidal suspensions~\cite{padding2004hab}. In the case of viral capsids, productive multimer-multimer bindings occur when aggregates' exposed contacts meet; changing the number of contacts per subunit (by mutation, for instance), thereby changing the sticky surface presented by a capsid of a given size, might be used to affect the importance of multimer binding to assembly pathways.

\section{Acknowledgements}

We thank Ludovic Berthier, Patrick Charbonneau and Thomas Ouldridge for correspondence. SW was supported initially by the US Department of Energy and subsequently by the BioSim European Union Network of Excellence, and acknowledges a Royal Society conference grant that made possible a collaborative visit. Computing facilities were provided in part by the Centre for Scientific Computing at the University of Warwick with support from the Science Research Investment Fund. EHF thanks the Miller Institute of Basic Research in Science for financial support.  MFH acknowledges funding from the HHMI-NIBIB Interfaces Initiative grant to Brandeis University, and Brandeis University startup funds. PLG acknowledges funding from the US Department of Energy.

\section{Appendix A}

The virtual-move scheme discussed in Ref~\cite{vmmc} and the main text accounts for the asymmetry of link formation at the level of the acceptance rate. Here we present a modification of this scheme in which link formation is forbidden if the corresponding link would not form under the reverse move. We find that the acceptance rate for collective motion is made simpler. Consider first a generic dynamic pseudocluster-formation procedure, moving from state $\mu$ to state $\nu$, in which a link is formed between particles $i$ and $j$ with probability $p_{ij}(\mu \to \nu)$. This probability is computed by making a virtual move of $i$. We subsequently specialize to the particular choice of $p_{ij}(\mu \to \nu)$ given in Equation~(\ref{virtual_link}).

\begin{enumerate}

\item Start in state $\mu$. Choose a seed particle, say $i$, and a move map. Add $i$ to the pseudocluster, the list of particles to be moved.
\item Choose a neighbor of $i$ not in the pseudocluster, and with which $i$ has not in the current move proposed a link. Call this neighbor $j$. With probability $p_{ij}(\mu \to \nu)$ (computed by moving $i$ under its virtual map) form a `pre-link' between $i$ and $j$ (not an automatic link, as described in the main text).
\begin{itemize}
\item If the pre-link does not form, we consider that a link has failed to form (outright failure). Choose another neighbor of $i$, say $k$, and return to stage 2, with the replacement $j \to k$.
\item If the pre-link $ij$ forms, compute the reverse link balance factor $f_{\rm reverse} = \min \left[ 1, \frac{p_{ij}(\nu \to \mu) }{p_{ij}(\mu \to \nu)} \right]$. 
\begin{itemize}
\item With probability $f_{\rm reverse}$, form a full link between $i$ and $j$. Add $j$ to the pseudocluster. Go to particle $j$ and proceed from step 2, with the replacement $j \to j'$ and $i \to j$. 
\item If a full link fails to form, record the link $ij$ as {\em frustrated}. Do not add $j$ to the pseudocluster. Choose another neighbor of $i$, say $k$, and return to stage 2, with the replacement $j \to k$.
\end{itemize}
\end{itemize}
\item Proceed until no more links remain to be tested, and evaluate the acceptance probability for the move.
\end{enumerate}
\noindent
The acceptance probability for the revised algorithm follows from a modification of Equation~(\ref{factor3}) in the main text. We choose to balance the total rates for forward and reverse moves involving a given realization ${\cal R}$ of internal pseudocluster links, and a realization of internal failed links `blind' to the nature of those failed links (whether frustrated or outright failed). The acceptance rate for the dynamic linking procedure described here, for a generic choice of $p_{ij}(\mu \to \nu)$, is
\bea
\label{factor1}
W_{\rm acc}^{\left(\mu \to \nu|{\cal R}\right)}&=& \Theta \left(n_{\rm c} -n_{{\cal C}} \right) \cal{D}(\cal{C}) \nonumber \\
&\times&\min \left\{ 1, {\rm e}^{-\beta(E_{\nu}-E_{\mu})} \nonumber \right. \\ 
&\times& \left. \frac{\prod_{\nu \to \mu}^{{\rm ext.}} q_{ij}(\nu \to \mu)}{\prod_{\mu \to \nu}^{{\rm ext.}} q_{ij}(\mu \to \nu) } \frac{\prod_{\nu \to \mu}^{{\rm ext.}} \hat{q}_{ij}(\nu \to \mu)}{\prod_{\mu \to \nu}^{{\rm ext.}} \hat{q}_{ij}(\mu \to \nu) } \nonumber \right. \\ 
&\times& \left. 
\frac{\prod_{\nu \to \mu}^{{\rm int.}} \tilde{q}_{ij}(\nu \to \mu)}{\prod_{\mu \to \nu}^{{\rm int.}} \tilde{q}_{ij}(\mu \to \nu) } \nonumber \right. \\
\hspace{-1cm} &\times& \left. 
\prod^{{\cal R}}_{\langle i j \rangle_{\ell}} \frac{p_{ij}( \nu \to \mu) \min \left(1,\frac{p_{ij}( \mu \to \nu)}{p_{ij}( \nu \to \mu)}\right)}{p_{ij}( \mu \to \nu) \min \left(1, \frac{p_{ij}( \nu \to \mu)}{p_{ij}( \mu \to \nu)}\right)} \right\}.
\eea
The first two lines of Equation~(\ref{factor1}) are as Equation~(\ref{factor3}) of the main text. Variables $q$ denote outright failed links between ${\cal C}$ and its environment. When we take $p_{ij}(\mu \to \nu)$ as in Equation~(\ref{virtual_link}) of the main text, such variables cancel the Boltzmann bond weights for all but a specialized class of moves (see main text). Variables $\hat{q}_{ij}(\mu \to \nu) =p_{ij}(\mu \to \nu) \left( 1-\min \left[ 1, \frac{p_{ij}(\nu \to \mu) }{p_{ij}(\mu \to \nu)} \right] \right)$ denote frustrated links between ${\cal C}$ and its environment. Frustrated links cannot form between the pseudocluster and its environment for both forward and reverse moves, and should such links form during the forward move then that move must be rejected. Variables $\tilde{q}$ denote unformed links internal to the pseudocluster, whether frustrated or outright rejected. We have that
\bea
\label{factor4}
\tilde{q}_{ij}(\mu \to \nu) &= &q_{ij}(\mu \to \nu) + \hat{q}_{ij}(\mu \to \nu) \nonumber \\
&=& 1-p_{ij}(\mu \to \nu) \nonumber \\
&+& p_{ij}(\mu \to \nu) \left( 1-\min \left[ 1, \frac{p_{ij}(\nu \to \mu) }{p_{ij}(\mu \to \nu)} \right] \right) \nonumber \\
&=& 1-p_{ij}(\mu \to \nu) \min \left[ 1, \frac{p_{ij}(\nu \to \mu) }{p_{ij}(\mu \to \nu)} \right],
\eea
and so $\tilde{q}_{ij}(\mu \to \nu)/\tilde{q}_{ij}(\nu \to \mu)$=1. Lastly, the product in the final line of Equation~(\ref{factor1}) runs over all fully-formed links. By construction of the linking procedure each quotient in this product is unity: we ensure that links formed during the forward move can also form during the reverse move.

If we take $p_{ij}(\mu \to \nu)$ as in Equation~(\ref{virtual_link}) of the main text, than the acceptance rate for the collective move is
\bea
\label{accept2}
W_{\rm acc}^{(\mu \to \nu|\cal{R})} &=&  \Theta \left(n_{\rm c} -n_{{\cal C}} \right) \cal{D}(\cal{C}) \nonumber \\
&\times& \min \left\{1,  \delta_{{\rm f.e.}} \prod_{\langle i j \rangle_{{\rm n}\leftrightarrow {\rm o}}} e^{-\beta\left( \epsilon_{ij}^{(\nu)} -\epsilon_{ij}^{(\mu)} \right)} \right\}.
\eea
The link-formation and link-failure factors internal to the pseudocluster have canceled. The factor $\delta_{\rm f.e.}$ in the third line of Equation~(\ref{accept2}) is unity if no frustrated links join the pseudocluster to its environment, and zero otherwise. This factor is required because a frustrated link indicates that a move of one particle relative to another has effectively been rejected. If we then do {\em not} form the link, in order to see if both particles are incorporated into the pseudocluster, we must reject any move in which both particles do {\em not} end up in the pseudocluster. The advantage of this scheme relative to that presented in the main text is that here links internal to the pseudocluster that form under the forward move, but do so with zero probability under the reverse move, do not automatically result in the rejection of that move.

\bibliography{bib}

\end{document}